\documentclass[aps,prd,preprint,a4paper,showpacs,nofootinbib,superscriptaddress]{revtex4-2}
\usepackage{bm}
\usepackage{indentfirst}
\usepackage{amsmath}
\usepackage{graphicx}
\usepackage{float}
\usepackage{amssymb}
\usepackage{subfigure}
\usepackage{amssymb}
\usepackage{hyperref}
\usepackage{epstopdf}
\usepackage[section]{placeins}

\usepackage[utf8]{inputenc}
\hypersetup{
    colorlinks=true,
    linkcolor=red,
    citecolor=blue,
}
\usepackage{color}
\usepackage[T1]{fontenc}
\usepackage{txfonts}
\usepackage{orcidlink}

\usepackage{adjustbox,lipsum}

\newcommand{\Rmnum}[1]{\expandafter\@slowromancap\romannumeral #1@} 
\newcommand{\bq}{\begin{equation}}
\newcommand{\eq}{\end{equation}}
\newcommand{\bqn}{\begin{eqnarray}}
\newcommand{\eqn}{\end{eqnarray}}
\newcommand{\nb}{\nonumber}
\newcommand{\lb}{\label}

\DeclareMathOperator{\sech}{sech}
\DeclareMathOperator\arctanh{arctanh}
\makeatother

\begin{document}

\title{On bifurcation and spectral instability of asymptotic quasinormal modes in the modified P\"oschl-Teller effective potential}

\author{Guan-Ru Li}
\affiliation{Faculdade de Engenharia de Guaratinguet\'a, Universidade Estadual Paulista, 12516-410, Guaratinguet\'a, SP, Brazil}

\author{Wei-Liang Qian}
\email{wlqian@usp.br}
\affiliation{Escola de Engenharia de Lorena, Universidade de S\~ao Paulo, 12602-810, Lorena, SP, Brazil}
\affiliation{Faculdade de Engenharia de Guaratinguet\'a, Universidade Estadual Paulista, 12516-410, Guaratinguet\'a, SP, Brazil}
\affiliation{Center for Gravitation and Cosmology, College of Physical Science and Technology, Yangzhou University, Yangzhou 225009, China}

\author{Ramin G. Daghigh}
\affiliation{Natural Sciences Department, Metropolitan State University, Saint Paul, Minnesota, 55106, USA}

\begin{abstract}
The P\"ochl-Teller effective potential mimics an asymptotically de Sitter black hole bounded by an event horizon and a cosmological one. 
Owing to the benefit of being analytically soluble, the asymptotic quasinormal modes in the modified P\"oschl-Teller potential have been extensively explored in the literature by various authors, and the results bear distinct features.
Specifically, for small discontinuities placed at the potential's peak, Skakala and Visser showed that the resulting modes lie primarily along the imaginary frequency axis, in line with the numerical results encountered for most black hole metrics.
However, it was also suggested that under ultraviolet perturbations, asymptotic modes are expected to lie parallel to the real axis, closely intervening with recent developments on spectral instability.
In this work, by numerical and semi-analytical approaches, we aim to resolve the above apparent ambiguity.
The numerical scheme is based on an improved version of the matrix method, which is implemented in compactified hyperboloidal coordinates on the Chebyshev grid.
It is demonstrated that both asymptotic behaviors indeed agree with the numerical findings, which is somewhat to one's surprise.
This is further complemented by a few intriguing details.
Specifically, we report the emergence of a novel branch of purely imaginary modes originating from a bifurcation in the asymptotic quasinormal mode spectrum.
Moreover, we demonstrate how the bifurcation and asymptotic modes evolve as the discontinuity moves away from the potential's peak, furnishing a dynamic picture as the spectral instability unfolds.
It is further argued that they can be partly attributed to the observed parity-dependent deviations occurring for the low-lying perturbed modes of the original P\"oschl-Teller effective potential.
Our findings are confirmed by independent numerical verifications.
Besides, they are reinforced by semi-analytic approaches, where one introduces a refined version of the approximation.
We discuss the implications of the obtained results.
\end{abstract}

\date{June 15th, 2024}

\maketitle


\newpage
\section{Introduction}\label{sec1}

The black hole, embodying one of the most intriguing phenomena in theoretical physics, epitomizes the extreme manifestations of gravity.
The successful detection of gravitational waves by the ground-based LIGO and Virgo collaboration~\cite{agr-LIGO-01, agr-LIGO-02, agr-LIGO-03, agr-LIGO-04} heralded the onset of a new era in observational astrophysics.
The achievement has further inspired many ongoing spaceborne projects LISA~\cite{agr-LISA-01}, TianQin~\cite{agr-TianQin-01, agr-TianQin-Taiji-review-01}, and Taiji~\cite{agr-Taiji-01}, with the potential for direct observation of ring down waveforms with desirable signal-to-noise ratio~\cite{agr-LISA-19, agr-LISA-20, agr-TianQin-05}.

Such specific waveforms mainly consist of quasinormal modes (QNMs)~\cite{agr-qnm-review-02, agr-qnm-review-03, agr-qnm-review-06}.
The physical interest arises from the fact that they are uniquely dictated by the underlying spacetime. As a result, their observation furnishes unambiguous information on the properties of spacetime in the vicinity of the horizon.
It was pointed out by Leaver~\cite{agr-qnm-21, agr-qnm-29} that QNMs are associated with the poles pertaining to Green's function of the underlying master equation, while the late-time tail is attributed to the branch cut, which usually follows a form that drops off as an inverse power in time~\cite{agr-qnm-tail-01}.

The properties of QNMs are closely associated with the notion of spectral instability.
The concept was pioneered by Nollert and Price~\cite{agr-qnm-35, agr-qnm-36}, as they demonstrated that minor perturbations, expressed as step functions, significantly impact high-overtone modes in the QNM spectrum. 
This demonstrated an unexpected instability of the QNM spectrum against ``ultraviolet'', namely, small-scale perturbations, challenging the assumption that a reasonable approximation of the effective potential ensures minimal deviation in the resulting QNMs. 
In~\cite{agr-qnm-50, agr-qnm-lq-03}, we further argued that even in the presence of discontinuity of a more moderate fashion, the asymptotic behavior of the QNM spectrum would be non-perturbatively modified. 
Specifically, high-overtone modes would shift along the frequency real axis instead of ascending the imaginary frequency axis observed for most black hole metrics~\cite{agr-qnm-continued-fraction-12, agr-qnm-continued-fraction-23}. 
Specifically, it was shown~\cite{agr-qnm-lq-03} that this phenomenon persists regardless of the discontinuity's distance from the horizon or its size.
Proposing the concept of spectral stability, Jaramillo {\it et al.}~\cite{agr-qnm-instability-07} analyzed the problem in the context of randomized perturbations to the metric. 
Their analysis revealed that the boundary of the pseudospectrum moves toward the real frequency axis. 
These results reinforce the universal instability of high-overtone modes triggered by ultraviolet perturbations.
The idea has been pushed forward by recent studies~\cite{agr-qnm-instability-08, agr-qnm-instability-13, agr-qnm-instability-14, agr-qnm-instability-19, agr-qnm-instability-26, agr-qnm-instability-32} and notably Cheung {\it et al.}~\cite{agr-qnm-instability-15} pointed out that even the fundamental mode can be destabilized under generic perturbations.

Spectral instability might have profound implications for observational astrophysics.
On the one hand, it is intrinsically intervened with the viability of black hole spectroscopy~\cite{agr-bh-spectroscopy-05, agr-bh-spectroscopy-06, agr-bh-spectroscopy-15, agr-bh-spectroscopy-18, agr-bh-spectroscopy-20, agr-bh-spectroscopy-36}.
As for real-world scenarios, gravitational radiation sources, such as black holes or neutron stars, are not isolated but interact with the surrounding matter. 
In other words, spacetime inevitably deviates from an ideally symmetric metric, even insignificantly, causing the emitted gravitational waves of the underlying QNMs to differ substantially from those of a pristine, isolated, compact object. 
This phenomenon has steered investigations towards the study of ``dirty'' black holes, explored by many authors~\cite{agr-bh-thermodynamics-12, agr-qnm-33, agr-qnm-34, agr-qnm-54}, opening new avenues in black hole perturbation theory.
On the other hand, the asymptotic modes that lie almost parallel to the real axis are closely related to the notion of echoes, an intriguing concept on late-stage ringing waveforms first proposed by Cardoso {\it et al.}~\cite{agr-qnm-echoes-01}. 
As a potential observable that might tell apart different but otherwise similar gravitational systems via their distinct properties near the horizon, the idea has incited many studies of echoes in various systems.
The ballpark of the latter embraces exotic compact objects such as gravastar~\cite{agr-eco-gravastar-02, agr-eco-gravastar-03}, boson star~\cite{agr-eco-gravastar-07}, and wormhole~\cite{agr-wormhole-01, agr-wormhole-02, agr-wormhole-10, agr-wormhole-11}. 
Like the late-time tail, echoes can also be attributed to the analytic properties of the Green's function.
Mark {\it et al.}~\cite{agr-qnm-echoes-15} evaluated Green's function in the frequency domain, and echoes are derived as a summation of a geometric series of products in reflection and transmission amplitudes.
In the context of Damour-Solodukhin type wormholes~\cite{agr-wormhole-12}, Bueno {\it et al.}~\cite{agr-qnm-echoes-16} explored the echoes by explicitly solving for specific frequencies when the transition matrix becomes singular.
In both cases, the emergence of echoes can be understood in terms of asymptotic QNMs~\cite{agr-qnm-echoes-20, agr-strong-lensing-correlator-15, agr-qnm-echoes-35}.
Specifically, the phenomenon is attributed to a spectrum of quasinormal frequencies lying uniformly along the real axis, where the interval between successive modes is related to the echo frequency.

The substantial implications of black hole spectral instability invite further investigations.
To our knowledge, explicit calculations regarding spectral instability have mainly been performed for randomized perturbations in effective potentials~\cite{agr-qnm-instability-07, agr-qnm-instability-13, agr-qnm-instability-14}.
Although being a simplification, the P\"oschl-Teller effective potential~\cite{agr-qnm-Poschl-Teller-01, agr-qnm-Poschl-Teller-02} mimics an asymptotically de Sitter black hole bounded by an event horizon and a cosmological one. 
Primarily because it is analytically soluble, it has been a worthy subject and extensively explored in the literature.
Existing results on spectral instability strongly indicate that the black hole QNM spectrum, particularly the high-overtone region of the spectrum, deforms and migrates toward the real frequency axis in response to even minuscule ultraviolet perturbations.
However, there is some potential ambiguity in relation to semi-analytical studies~\cite{agr-qnm-Poschl-Teller-03, agr-qnm-Poschl-Teller-04} carried out by Skakala and Visser {\it et al.}, which have indicated a somewhat different picture.
When modifying the P\"ochl-Teller effective potential by placing an insignificant discontinuity at the peak of the potential, which seemingly qualifies as a small-scale ultraviolent perturbation, the authors encountered the asymptotic QNMs with a significant imaginary part.
In fact, asymptotic modes were found to suffer only perturbative modifications compared to those for the original P\"oschl-Teller potential~\cite{agr-qnm-Poschl-Teller-01, agr-qnm-Poschl-Teller-02}.
In other words, they lie primarily along the imaginary frequency axis, reminiscent of most unperturbed black hole metrics.
In contrast, these results are qualitatively different when the perturbation is placed further away from the black hole.
As discussed in Ref.~\cite{agr-qnm-lq-03}, the QNMs should asymptotically lie parallel to the real axis for the latter case.
Therefore, it is not entirely clear whether some of the following statements are true.
Is there an undiscovered branch of asymptotic modes lying along the real axis when the discontinuity is placed at the origin? 
Do some asymptotic modes persist and still sit along the imaginary axis when the perturbations are further away from the origin?
As the discontinuity moves away from the horizon toward the spatial infinity, will the original asymptotic modes be significantly deformed and migrate to lay parallel to the real axis?
The answers to the above questions will likely impact our understanding of the spectral instability and potentially lead to observational implications on gravitational wave detection.

The present study aims to address some of the above considerations.
Given that the existing semi-analytic approaches can only cover a few limited scenarios, a more comprehensive study is desirable to bring us a better understanding of the collective behavior of QNMs under metric perturbations. 
By numerical and semi-analytical approaches, we aim to resolve the above apparent ambiguity.
The numerical scheme is based on an improved version of the matrix method, which is implemented in hyperboloidal coordinates~\cite{agr-qnm-hyperboloidal-01} on the Chebyshev grid~\cite{agr-qnm-lq-matrix-11} in the presence of discontinuity~\cite{agr-qnm-lq-matrix-06}.
We show that both asymptotic behaviors indeed agree with the numerical findings, bridged by a migration of the QNM spectrum as the discontinuity moves away from the maximum of the effective potential.
Somewhat to one's surprise, we also observe some intriguing details in the QNM spectrum, which, to our knowledge, have yet to be reported.
In particular, we report the emergence of a branch of purely imaginary modes in the presence of ultraviolet perturbations.
Such a branch originates from a bifurcation in the asymptotic QNM spectrum.
Regarding Skakala and Visser's results, we observe parity-dependent deviations occurring for the perturbed relatively low-lying modes of the original Pöschl-Teller effective potential.
Specifically, the deviations from the original QNMs depend explicitly on the spatial reflection symmetry of the waveforms.
It is understood that the bifurcation results from the spectral instability triggered by such parity-dependent corrections.
We show that the evolution of the QNM spectrum can be understood from a dynamical perspective as the spectral instability unfolds when the discontinuity in the effective potential moves towards spatial infinity.  
To verify our findings, we also perform independent numerical approaches and the results agree well with those obtained using the matrix method.
Moreover, these results are confirmed semi-analytically by introducing a refined approximation.
We discuss the implications of the obtained results.

The remainder of the paper is organized as follows.
In Sec.~\ref{sec2}, we review the existing analytic approaches and results on modified P\"oschl-Teller potential, from which we point out the drastic difference in the asymptotic QNMs.
Subsequently, in Sec.~\ref{sec3}, we elaborate on an improved version of the matrix method, which will be subsequently used as one of the independent numerical approaches for the problem. 
Compared to previous versions, the scheme is generalized and implemented in compactified hyperboloidal coordinates on the Chebyshev grid.
The numerical results are presented in Sec.~\ref{sec4}.
In Secs.~\ref{sec5} and~\ref{sec6}, we aim to reinforce our numerical findings by employing different approaches.
Inclusively, we elaborate on a semi-analytic analysis based on a refined version of the approximation.
The last section includes further discussions and concluding remarks.
We relegate the somewhat tedious analytical derivations to Appx.~\ref{appA} and~\ref{appB}.

\section{Asymptotic quasinormal modes in modified P\"oschl-Teller potential}\label{sec2}

In this section, we briefly revisit the formalism for QNMs and existing semi-analytic approaches and results on the modified P\"oschl-Teller potential, which furnishes the basis for the analysis carried out in the preceding sections.

The study of black hole perturbation theory often leads to exploring the solution of the radial part of the master equation~\cite{agr-qnm-review-03,agr-qnm-review-06},
\begin{eqnarray}
\frac{\partial^2}{\partial t^2}\Psi(t, r_*)+\left(-\frac{\partial^2}{\partial r_*^2}+V_\mathrm{eff}\right)\Psi(t, r_*)=0 ,
\label{master_eq_ns}
\end{eqnarray}
where the spatial coordinate $r_*$ is known as the tortoise coordinate, and the effective potential $V_\mathrm{eff}$ is governed by the given spacetime metric, spin ${\bar{s}}$, and angular momentum with multipole number $\ell$ of the waveform.
For instance, the Regge-Wheeler potential $V_\mathrm{RW}$ for the Schwarzschild black hole metric is
\bqn
V_\mathrm{eff} = V_\mathrm{RW}=F\left[\frac{\ell(\ell+1)}{r^2}+(1-{\bar{s}}^2)\frac{r_h}{r^3}\right],
\lb{Veff_RW}
\eqn
where 
\bqn
F=1-r_h/r ,
\lb{f_RW}
\eqn
and $r_h=2M$ is the event horizon radius determined by the black hole's ADM mass, $M$.  
Here, we use geometric units with $G=c=1$.
The tortoise coordinate $r_*\in(-\infty,+\infty)$ is related to the radial coordinate $r\in [0,+\infty)$ by the relation $r_*=\int dr/F(r)$.

The black hole QNMs are determined by solving the eigenvalue problem defined by Eq.~\eqref{master_eq_ns} in the frequency domain
\begin{equation}
\frac{d^2\Psi(\omega, r_*)}{dr_*^2}+[\omega^2-V_\mathrm{eff}]\Psi(\omega, r_*) = 0 , \label{master_frequency_domain}
\end{equation}
by the following boundary conditions in asymptotically flat spacetimes
\begin{equation}
\Psi \sim
\begin{cases}
   e^{-i\omega_{n} r_*}, &  r_* \to -\infty, \\
   e^{+i\omega_{n} r_*}, &  r_* \to +\infty,
\end{cases}
\label{master_bc0}
\end{equation}
which indicates an ingoing wave at the horizon and an outgoing wave at infinity.
The subscript $n$ represents the overtone number.
Apart from the initial burst, the temporal profile of the waveform $\Psi$ is characterized by the quasinormal oscillations and late-time tail.
The QNMs are governed by the eigenvalues $\omega_{n}$, known as the quasinormal frequencies.
They are typically complex numbers attributed to the dissipative nature of Eq.~\eqref{master_bc0}.

The modified P\"oschl-Teller potential explored extensively in the literature possesses the following form
\bqn
V_\mathrm{eff} 
= \begin{cases}
   V_\mathrm{PT}(r_{*}; V_{0-},b_-), &  r_* < r_{*c}, \\
   V_\mathrm{PT}(r_{*}; V_{0+},b_+), &  r_* > r_{*c}, 
\end{cases},
\lb{Veff_MPT}
\eqn
where $r_{*c}$ is a point of discontinuity and the P\"ochl-Teller potential is given by
\bqn
 V_\mathrm{PT}(r_{*}; V_0,b)={V_0}~{\sech}^2\left(\frac{r_{*}}{b}\right),
\lb{potential_PT}
\eqn
where $V_0$ and $b$ are two parameters governing the shape of the potential.

It is worthwhile to note that, in the vicinity of the horizon, a typical black hole effective potential, such as Eq.~\eqref{Veff_RW}, is suppressed by an exponential form when expressed as a function of tortoise coordinate, namely,
\bqn
V_\mathrm{eff} \sim \exp\left(-2\kappa r_*\right) ,
\eqn
where $\kappa$ is the surface gravity of the black hole. 
In this regard, the scenario given by the P\"ochl-Teller potential Eq.~\eqref{potential_PT} can be viewed as an approximation to asymptotically de Sitter black holes, where the region of interest is bounded by an event horizon and a cosmological one.

In Refs.~\cite{agr-qnm-Poschl-Teller-03, agr-qnm-Poschl-Teller-04}, the authors consider the case where $r_{*c}=0$ and $b_+\simeq b_-$.
An analytic approach was carried out by focusing on the highly damped asymptotic modes that possess significant imaginary parts.
Moreover, by using a particular differential identity for the hypergeometric function, Bailey's theorem, and the asymptotic expansion of the Gamma function, the QNMs can be identified by solving a trigonometric equation
\bqn
\cos\left(-\pi i \omega\Delta\right)-{\cos}\left(-2\pi i {\omega}b_*\right)=2\cos(\pi\alpha_{-})\cos(\pi\alpha_{+}) ,
\lb{Visser_QNF_condition}
\eqn
where
\bqn
\alpha_{\pm}=\sqrt{\frac{1}{4}-V_{0\pm}b_{\pm}^2} , \lb{Visser_parameters}
\eqn
and one has introduced the definitions 
\bqn
b_* &=& \frac12\left(b_{-}+b_{+}\right) ,\nb\\
\Delta &=& b_{-}-b_{+} = 2\epsilon b_* . \lb{Visser_parameters2}
\eqn
Regarding the scenario relevant to black hole metric perturbations, one considers the case $b_+\simeq b_-$, and therefore, $\epsilon$ (or $\Delta$) is treated as an insignificant quantity.
At the limit $\epsilon\to 0$, we can ignore the modulation of the term $\cos\left(\pi i \omega\Delta\right)$ in Eq.~\eqref{Visser_QNF_condition} and write down the zeroth order result\footnote{In this paper, we adopt the convention that $\Im\omega_n < 0$ in accordance with Eq.~\eqref{master_bc0}.
Therefore, the resulting expressions differ from those in~\cite{agr-qnm-Poschl-Teller-03, agr-qnm-Poschl-Teller-04}.}
\bqn
\omega^{(0)}_n=i\frac{\arccos\left(1-2\cos(\pi\alpha_{-})\cos(\pi\alpha_{+})\right)}{2\pi b_*}-i\frac{n}{b_*} .
\lb{Visser_result0}
\eqn
In particular, when $b_{-}=b_{+}=b$ and $V_{0-}=V_{0+}=V_0$ one has $\alpha_+=\alpha_-=\alpha$ and Eq.~\eqref{Visser_result0} readily falls back to 
\bqn
{\omega}^{(0)}_n=\omega_n^\mathrm{PT}=i\frac{1}{b}\left(-n-\frac{1}{2}\pm\alpha\right),
\lb{qnm_PT}
\eqn
namely, the result of the original P\"ochl-Teller potential\footnote{Here, the derivation is carried out for $n\gg 1$. By comparing against the original P\"ochl-Teller potential's QNMs, it must be a positive integer.}~\cite{agr-qnm-Poschl-Teller-01, agr-qnm-Poschl-Teller-02}.

To take into account the first-order contribution, one retains the relevant terms in deriving Eq.~\eqref{Visser_result0} and finds~\cite{agr-qnm-Poschl-Teller-03}
\bqn
\omega^{(1)}_n &=& i\frac{\arccos\left[\cos(i\pi{\omega}^{(0)}_n\Delta)-2{\cos(\pi\alpha_{-})}{\cos(\pi\alpha_{+})}\right]}{2\pi b_*}-i\frac{n}{b_*} \nb\\
&=& i\frac{\arccos\left[\cos(i\pi{\omega}^{(0)}_n\epsilon b_*)+\cos(i2\pi{\omega}^{(0)}_nb_*)-1\right]}{2\pi b_*}-i\frac{n}{b_*},
\lb{Visser_result1}
\eqn
where the second line is obtained by substituting the zeroth-order result, Eq.~\eqref{Visser_result0}.
The above results indicate that the asymptotic QNMs are largely parallel to the imaginary axis.
This is a typical characteristic of most known black hole metrics.
In other words, the results of~\cite{agr-qnm-Poschl-Teller-03, agr-qnm-Poschl-Teller-04} indicate that the asymptotic behavior of the black hole QNM spectrum largely remains unchanged under perturbations placed at the origin.
We note that the first line of Eq.~\eqref{Visser_result1} can be used as an iterative relation for higher-order results.
We relegate further discussions to Sec.~\ref{sec5}.

It is not straightforward to generalize the above semi-analytic approach to the case when the junction is not at the origin. 
However, the asymptotic behavior of QNMs has been analyzed when the discontinuity is placed further away from the maximum of the effective potential.
In~\cite{agr-qnm-lq-03}, some of us explored a scenario where the P\"ochl-Teller effective potential is truncated at the point $r_{*c}\gg 1$.
At such a limit, one starts with the assumption that the asymptotic mode has a significant real part.
In practice, QNMs can be evaluated by demanding that the Wronskian of the underlying Green's function, which can be conveniently evaluated at the truncation point, vanish.
As it turns out, the originally outgoing wave at $r_*\to\infty$ will receive an insignificant fraction of the ingoing wave, which pushes the quasinormal frequencies away from the original poles of the Gamma function. 
It was found that the resulting QNMs have the following asymptotic behavior
\bqn
\omega_n = \frac{\pi n}{r_{*0}} -i\left[\frac{\ln(\pi n)}{r_{*0}}+\frac{1}{b_-}-\frac{\ln r_{*0}}{r_{*0}}-\frac{\ln V_{0-}}{2r_{*0}}\right]+ O(1).
\lb{PT_echo_modes}
\eqn
As mentioned above, it is not clear whether the asymptotic modes dictated by Eq.~\eqref{Visser_result1} persist as the discontinuity moves away from the origin or, as implied by the spectral instability analysis, they will be significantly deformed to predominantly lay parallel to the real axis, per Eq.~\eqref{PT_echo_modes}. 
In what follows, we elaborate on the matrix method, which is subsequently employed to evaluate the high-overtone numerically.

\section{An improved matrix method implemented on Chebyshev grids in hyperboloidal coordinates}\label{sec3}

The {\it matrix method}~\cite{agr-qnm-lq-matrix-01} is inspired by one of the most accurate numerical approaches to date for evaluating the QNMs, the continued fraction method~\cite{agr-qnm-continued-fraction-01, agr-qnm-continued-fraction-03, agr-qnm-continued-fraction-04}.
While appropriately taking into account the asymptotic waveform, such an approach expands the waveform at a given coordinate and reiterates the master equation~\eqref{master_eq_ns} as an iterative relation between the expansion coefficients, which can be equivalently expressed in a mostly diagonal matrix form.
Thus, the QNMs problem is effectively solved by using Hill's determinant.
Following this line of thought, instead of a given position, one may discretize the entire spatial domain and perform the expansions of the waveform on the entire grid~\cite{agr-qnm-lq-matrix-01}.
Subsequently, the master equation can be formulated into a mostly dense matrix equation, and the QNM problem is reiterated as an algebraic nonlinear equation for the complex frequencies.
Such an approach was employed to evaluate QNMs of compact stars~\cite{agr-qnm-star-27, agr-qnm-star-34} pioneered by Kokkotas, Ruoff, Boutloukos, and Nollert and recently promoted by Jansen~\cite{agr-qnm-59}.
Some of us have further pursued the idea and developed its applications in the context of black hole perturbation theory. 
Besides the metrics with spherical symmetry~\cite{agr-qnm-lq-matrix-02}, the matrix method can be applied to black hole spacetimes with axial symmetry~\cite{agr-qnm-lq-matrix-03} and a system composed of coupled degrees of freedom~\cite{agr-qnm-lq-matrix-07}.
The approach is shown to be effective in dealing with different boundary conditions~\cite{agr-qnm-lq-matrix-04} and dynamic black holes~\cite{agr-qnm-lq-matrix-05}.
More recently, the original method was generalized to handle effective potentials containing discontinuity~\cite{agr-qnm-lq-matrix-06} and pushed to higher orders~\cite{agr-qnm-lq-matrix-08}, inclusively implemented on Chebyshev grids~\cite{agr-qnm-lq-matrix-11}.
The latter is related to the notorious drawback of the equispaced interpolation points.
Specifically, polynomial interpolation based on a uniform grid is often liable to Runge's phenomenon, characterized by significant oscillations at the edges of the relevant interval, closely related to an increasing Lebesgue constant~\cite{book-approximation-theory-Rivlin, agr-qnm-lq-matrix-10}.
In other words, even though the uniform convergence over the interval in question is provided, an interpolation of a higher degree does not necessarily guarantee an improved accuracy, similar to the Gibbs phenomenon in Fourier series approximations.
On top of the above recipe, there is one more crucial ingredient, namely, the hyperboloidal coordinates~\cite{agr-qnm-hyperboloidal-01}.
In the place of spatial infinity adopted by most approaches as the boundary, the hyperboloidal coordinates guarantee that the boundary condition of the problem is evaluated at null infinity, a physically adequate and numerically favorable choice.
As shown shortly below, the improved matrix method can precisely evaluate high overtones while capturing some fine details not reported in the literature.

In implementing the matrix method, one first rewrites the master equation Eq.~\eqref{master_eq_ns} from $(t, r_*)$ into the compactified hyperboloidal coordinates $(\tau, x)$, where $x\in [-1,1]$~\cite{agr-qnm-hyperboloidal-01}.
Following Ref.~\cite{agr-qnm-instability-07}, it is convenient to first transform the coordinates into dimensionless quantities $(\overline{t}, \overline{x})$ by introducing a length scale $\lambda$. 
Specifically,
\bqn
\overline{t}=\frac{t}{\lambda},~~~~\overline{x}=\frac{r_{*}}{\lambda},~~~~{\hat{V}_\mathrm{eff}=\lambda^2 V_\mathrm{eff}},
\lb{dimensionless_quantities}
\eqn
where the choice of $\lambda$ is rather arbitrary, primarily aimed to simplify the resultant master equation. 
Subsequently, the compactified hyperboloidal coordinates $(\tau,x)$ are defined by
\bqn
\overline{t}=\tau-H(\overline{x}),~~~~\overline{x}=G(x).
\lb{compactified_hyperboloidal_approach}
\eqn
where the function $G(x)$ introduces a spatial compactification while the height function $H(x)$ is defined to guarantee that the boundary at $x=\pm 1$ for a given $\tau$ is a null infinity.
Specifically, 
\bqn
\partial_xH &\le& 1 ,\nb\\
\lim\limits_{x\to \pm 1} \partial_xH &=& \pm 1,
\lb{H_relation}
\eqn
where
$\partial_xH \equiv \frac{\partial H\left(G(x)\right)}{\partial x}$.
In contrast, the original boundary of the master equation Eq.~\eqref{master_eq_ns} is placed at the spatial infinity $\overline{x}=\pm\infty$ at a given $\overline{t}$, or equivalently $r_* = \pm\infty$ at a given $t$.
As initially pointed out by Zenginoğlu~\cite{agr-qnm-hyperboloidal-01}, this significantly simplifies the boundary conditions since any localized initial perturbations will never traverse the boundary owing to the causality.
Subsequently, one is free from the somewhat cumbersome outgoing wave boundary conditions Eq.~\eqref{master_bc0} defined at the spacelike infinity. 
It is not difficult to show that the asymptotic geometry of the $\tau$-hypersurface after such a transformation is a hyperboloid, after which the transformed coordinates are named. 
Under coordinates $(\tau,x)$, the master equation Eq.~\eqref{master_eq_ns} possesses the following form:
\bqn
\left[\left(1-\left(\frac{\partial_{x}H}{\partial_{x}G}\right)^2\right)\partial^2_{\tau}-2\frac{\partial_{x}H}{~(\partial_{x}G)^2}\partial_{\tau}\partial_{x}-\frac{1}{\partial_{x}G}\partial_{x}\left(\frac{\partial_{x}H}{\partial_{x}G}\right)\partial_{\tau}-\frac{1}{\partial_{x}G}\partial_{x}\left(\frac{1}{\partial_{x}G}\partial_{x}\right)+\hat{V}_\mathrm{eff}\right]\Psi=0 .
\lb{transformed_master_equation}
\eqn
By plugging the separation of the variables
\bqn
\Psi(\tau,x)=e^{-i\omega \lambda \tau}\psi(x),
\lb{wave_function}
\eqn
into the master equation, one finds the counterpart of Eq.~\eqref{master_frequency_domain}
\bqn
\left[a(\omega,x)+b(\omega,x)\partial_{x}+c(\omega,x)\partial_{xx}\right]\psi=0,
\lb{simplified_master_equation}
\eqn
where
\bqn
a(\omega,x) &=& \lambda\omega\left(\frac{i(\partial_{x}H)(\partial_{xx}G)}{(\partial_{x}G)^3}+\frac{\lambda\omega(\partial_{x}H)^2-i(\partial_{xx}H)}{(\partial_{x}G)^2}-\lambda\omega\right)+\hat{V}_\mathrm{eff},\nb\\
b(\omega,x) &=& \frac{(\partial_{xx}G)-2i\lambda\omega(\partial_{x}G)(\partial_{x}H)}{(\partial_{x}G)^3},\nb\\
c(\omega,x) &=& -\frac{1}{(\partial_{x}G)^2},
\eqn
which is an ordinary second-order differential equation in $x$ defined on the interval $[-1,1]$.
It is readily to show that the frequency $\omega$ in Eq.~\eqref{simplified_master_equation} is identical to that defined in Eq.~\eqref{master_frequency_domain}, as the transformation leaves the timelike Killing field in the exterior domain invariant.

To proceed, we discretize Eq.~\eqref{simplified_master_equation} on a Chebyshev grid~\cite{book-approximation-theory-Rivlin} of $N$ nodes to find a matrix equation~\cite{agr-qnm-lq-matrix-11}
\bqn
\mathcal{M(\omega)}\Vec{\psi}=0,
\lb{shorthand_master_equation}
\eqn
where the coefficient $\mathcal{M(\omega)}$ is an $N\times N$ matrix whose elements are functions of the frequency and $\vec{\psi}$ is a $1\times N$ column matrix associated with the waveform,
\bqn
\vec{\psi}=\left(\psi_0,\psi_1,\cdots,\psi_N\right)^{\rm T} ,
\lb{origin_psi}
\eqn
where $\psi_i\equiv\psi(x_i)$ are evaluated on the Chebyshev grid
\bqn
x_i={\cos}\left(\frac{i}{N}\pi\right),
\lb{Chebyshev_type_grids}
\eqn
where $i=0,\cdots,N$.

Subsequently, the quasinormal frequencies are attained at vanishing determinant
\bqn
\rm det(\mathcal{M(\omega)})=0.
\lb{origin_solve}
\eqn

In principle, Eq.~\eqref{origin_solve} can be solved using the standard numerical methods. 
In~\cite{agr-qnm-59}, Jansen proposed a rather effective algorithm when the power in frequency is not significant.
The implementation is based on the fact that, in practice, the elements of the coefficient matrix $\mathcal{M(\omega)}$ are polynomials in $\omega$ of finite order.
Specifically, this allows us to reiterate the matrix equation in the form
\bqn
\mathcal{M(\omega)}\Vec{\psi}=\left[M_0+M_1\omega+\cdots+M_k\omega^k\right]\vec{\psi}=0 ,\lb{mJason}
\eqn
where we have assumed that the highest power in all the polynomials is $k$.
It is straightforward to show that Eq.~\eqref{mJason} can be formally written as
\bqn
\left[\tilde{\mathcal{M}}_0-\omega \tilde{\mathcal{M}}_1\right]\vec{\psi}_{\omega}=0,
\lb{Aron_form_approach}
\eqn
where
\bqn
\tilde{\mathcal{M}}_0=
\begin{pmatrix}
M_0&M_1&M_2&\cdots&M_{k-1}\\
0&1&0&\cdots&0\\
0&0&1&\cdots&0\\
\vdots&\vdots&\vdots&\ddots&\vdots\\
0&0&0&\cdots&1
\end{pmatrix},
\tilde{\mathcal{M}}_1=\begin{pmatrix}
0&0&0&\cdots&0&-M_{k}\\
1&0&0&\cdots&0&0\\
0&1&0&\cdots&0&0\\
\vdots&\vdots&\vdots&\ddots&\vdots&\vdots\\
0&0&0&\cdots&1&0
\end{pmatrix},
\Vec{\psi}_{\omega}=
\begin{pmatrix}
\vec{\psi}\\
\omega^1\vec{\psi}\\
\omega^2\vec{\psi}\\
\vdots\\
\omega^{k-1}\vec{\psi}
\end{pmatrix},
\eqn
where $M_i$ and $\tilde{\mathcal{M}}_i$ are, respectively, $N\times N$ and $(Nk)\times (Nk)$ matrices.
The above equation gives rise to the following linear form
\bqn
\tilde{\mathcal{M}}_0=\omega \tilde{\mathcal{M}}_1 ,
\lb{Aron_form_solve}
\eqn
which can be readily solved using \textit{Mathematica}'s subroutines {\it Eigenvalue} or {\it Eigensystem}. 

Lastly, the discontinuity in the modified P\"oschl-Teller potential Eq.~\eqref{Veff_MPT} can be handled by replacing the corresponding row of the matrix by the Wronskian's vanishing condition evaluated at the discontinuous point~\cite{agr-qnm-lq-matrix-06}.

\section{Numerical results}\label{sec4}

In this section, we explore the QNMs of the modified P\"oschl-Teller effective potential from a numerical perspective.
In addition to primarily using the improved matrix method elaborated in the last section, we will also employ a complementary independent numerical scheme based on the evaluation of the Wronskian. 
Our findings mainly consist of the following results:
\begin{enumerate}
\item We numerically reproduce the semi-analytic results~\cite{agr-qnm-Poschl-Teller-03, agr-qnm-Poschl-Teller-04} on QNM when the P\"oschl-Teller effective potential is perturbed by a discontinuity placed at the origin $x_c=0$. 
\item At a higher precision, small deviations in the quasinormal frequency are observed and verified by an independent approach.
\item When a discontinuity is placed further away from the potential's peak~\cite{agr-qnm-lq-03}, we observe a bifurcation of the QNMs with one branch lying along the real axis, associated with the spectral instability~\cite{agr-qnm-instability-07}, and another branch that asymptotically consists of purely imaginary modes.
\item A dynamic picture bridges two types of distinct asymptotic QNMs as the spectral instability unfolds. 
Specifically, the bifurcation and asymptotic modes evolve as the discontinuity moves away from the potential's peak.
\end{enumerate}

We consider the effective potential given by Eq.~\eqref{Veff_MPT} and compare the semi-analytic results obtained for specific scenarios.
As discussed above in Sec.~\ref{sec2}, for the first scenario where $r_{*c}=0$ and $b_+\simeq b_-$, the asymptotic QNMs are given by Eq.~\eqref{Visser_result1}.
For the second scenario, one has $r_{*c}\gg 1$ and $V_{0+}=0$, and the asymptotic QNMs are governed by Eq.~\eqref{PT_echo_modes}.

By employing the modified matrix method, one has $V_\mathrm{eff}=V_\mathrm{PT}$. 
We choose the scaling
\bqn
\lambda = b,
\lb{choiceLambda}
\eqn
on either side of the discontinuity, and using Eq.~\eqref{dimensionless_quantities} we have
\bqn
\hat{V}_\mathrm{eff}=V(\overline{x})=
\left\{
    \begin{aligned}
    b_{-}^2V_{0-}{\sech}^2(\overline{x})~~~~\mathrm{for}~~~~\overline{x} < G(x_c)\\
    b_{+}^2V_{0+}{\sech}^2(\overline{x})~~~~\mathrm{for}~~~~\overline{x} > G(x_c)\\
    \end{aligned}
\right. ,
\lb{discontinuous_potential_PT}
\eqn
where $x_c$ is the discontinuity in the effective potential expressed in the Bizo\'n-Mach coordinates defined as
\bqn
H(x)&=&\frac{1}{2}{\ln}(1-x^2),\nb\\
G(x) &=& \arctanh(x) .
\lb{compactified_hyperboloidal_approach_PT}
\eqn
The resulting master equation on either side of the discontinuity reads
\bqn
(1-x^2)\left[(b^2\omega^2+ib\omega-b^2V_0)+2x(ib\omega-1)\partial_{x}+(1-x^2)\partial_{x,x}\right]\psi=0 ,
\lb{transformed_master_equation_PT}
\eqn
which further simplifies to
\bqn
\left[(b^2\omega^2+ib\omega-b^2V_0)+2x(ib\omega-1)\partial_{x}+(1-x^2)\partial_{x,x}\right]\psi=0 ,
\lb{simplified_master_equation_PT}
\eqn
where one notes substantial simplification in the form of the effective potential owing to the appropriate choice Eq.~\eqref{choiceLambda}, which implies ${\sech}^2(\overline{x})\to 1-x^2$.

Specifically, we have the following piecewise master equation
\bqn
{[}(b_{-}^2\omega^2+ib_{-}\omega-b_{-}^2V_{0-})+2x(ib_{-}\omega &-& 1)\partial_{x}+(1-x^2)\partial_{x,x}{]}\psi_{-} = 0,~~~~\mathrm{for}~~~~x\in (-1, x_c) ,\nonumber\\
{[}(b_{+}^2\omega^2+ib_{+}\omega-b_{+}^2V_{0+})+2x(ib_{+}\omega &-& 1)\partial_{x}+(1-x^2)\partial_{x,x}{]}\psi_{+} = 0,~~~~\mathrm{for}~~~~x\in (x_c, +1),
\lb{discontinuous_simplified_master_equation_PT}
\eqn
and the junction condition~\cite{agr-qnm-lq-matrix-06}
\bqn
\psi_{-}\partial_{x}\psi_{+} &-& \psi_{+}\partial_{x}\psi_{-} =0,~~~~~~~~~~~~~~~~~~~~~~~\mathrm{for}~~~~x=x_c.
\lb{junctionCondition}
\eqn

As discussed in the last section, the matrix method is implemented by discretizing and rewriting Eqs.~\eqref{discontinuous_simplified_master_equation_PT} into a matrix equation while appropriately substituting one row of the matrix by Eq.~\eqref{junctionCondition}.
To be more specific, the junction condition is implemented at the spatial coordinate, which corresponds simultaneously to the last row of the submatrix associated with the first line and the first row of the submatrix related to the second line of Eqs.~\eqref{discontinuous_simplified_master_equation_PT}.
Notably, one does not enforce the waveform to vanish at the boundary $x=\pm 1$~\cite{agr-qnm-hyperboloidal-01}.

\begin{figure}[htp]
\centering
\includegraphics[scale=0.65]{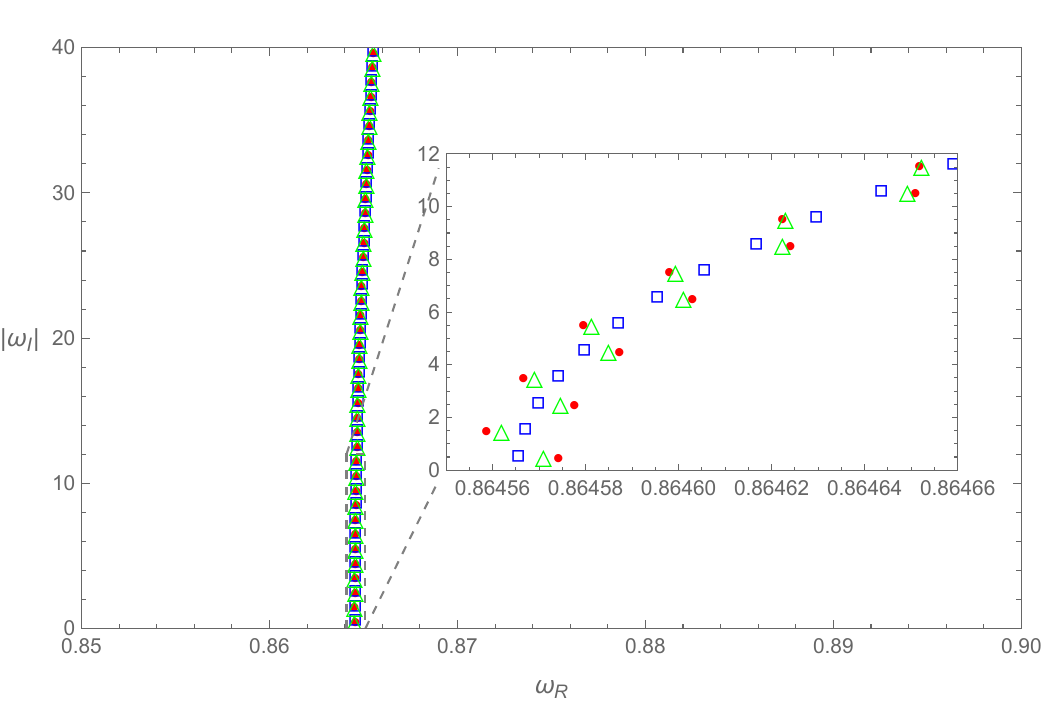}
\caption{Asymptotic QNMs for the modified P\"oschl-Teller effective potential Eq.~\eqref{Veff_MPT} evaluated using different approaches. 
The calculations are carried out using the metric parameters $x_c=0$, $(V_{0-}, b_{-}) = (1, 1)$, and $(V_{0+}, b_{+}) = (1, 0.99)$.
This corresponds to the scenario explored in~\cite{agr-qnm-Poschl-Teller-03, agr-qnm-Poschl-Teller-04} where a discontinuity is placed at the origin. 
The numerical results obtained by the matrix method discussed in Sec.~\ref{sec3} are shown in filled red circles, the semi-analytic results given in~\cite{agr-qnm-Poschl-Teller-03, agr-qnm-Poschl-Teller-04} are represented by empty blue squares, empty green triangles indicate the semi-analytic results derived in Sec.~\ref{sec5}.
The inset of the figure shows a zoomed-in section of a few lowest-lying modes indicated by the dashed square box.}
\label{fig1}
\end{figure}

\begin{figure}[htp]
\centering
\includegraphics[scale=0.65]{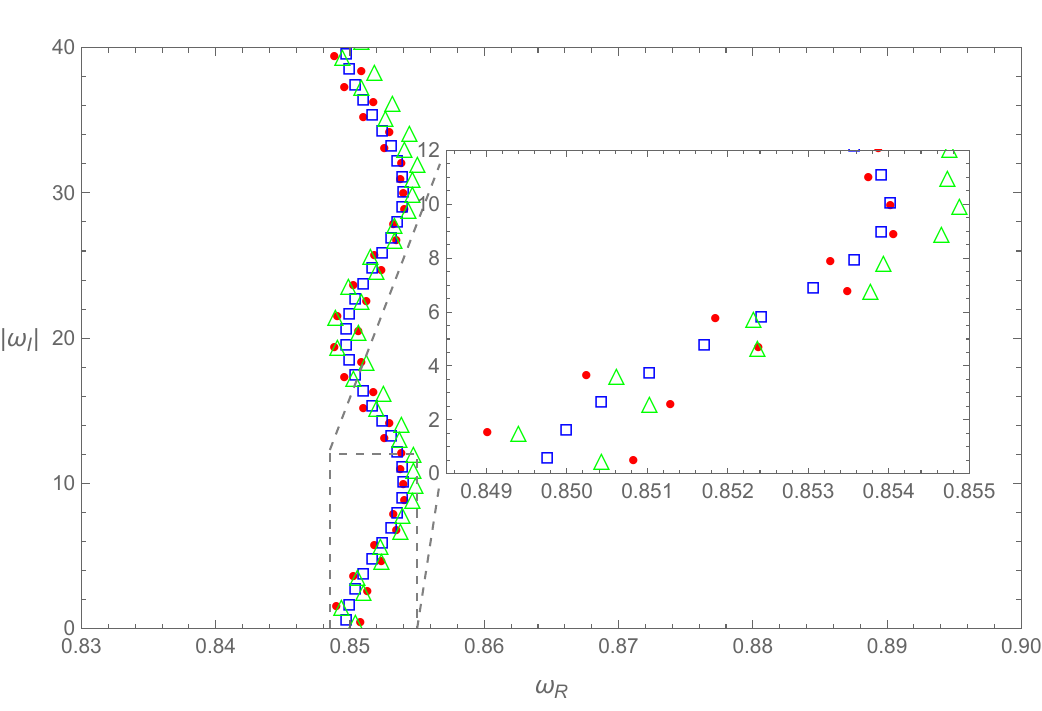}
\caption{The same as Fig.~\ref{fig1}, but the calculations are carried out using the metric parameters $x_c=0$, $(V_{0-}, b_{-}) = (1, 1)$, and $(V_{0+},b_{+})=(1,0.9)$.}
\label{fig2}
\end{figure}

\begin{figure}[htp]
\centering
\includegraphics[scale=0.65]{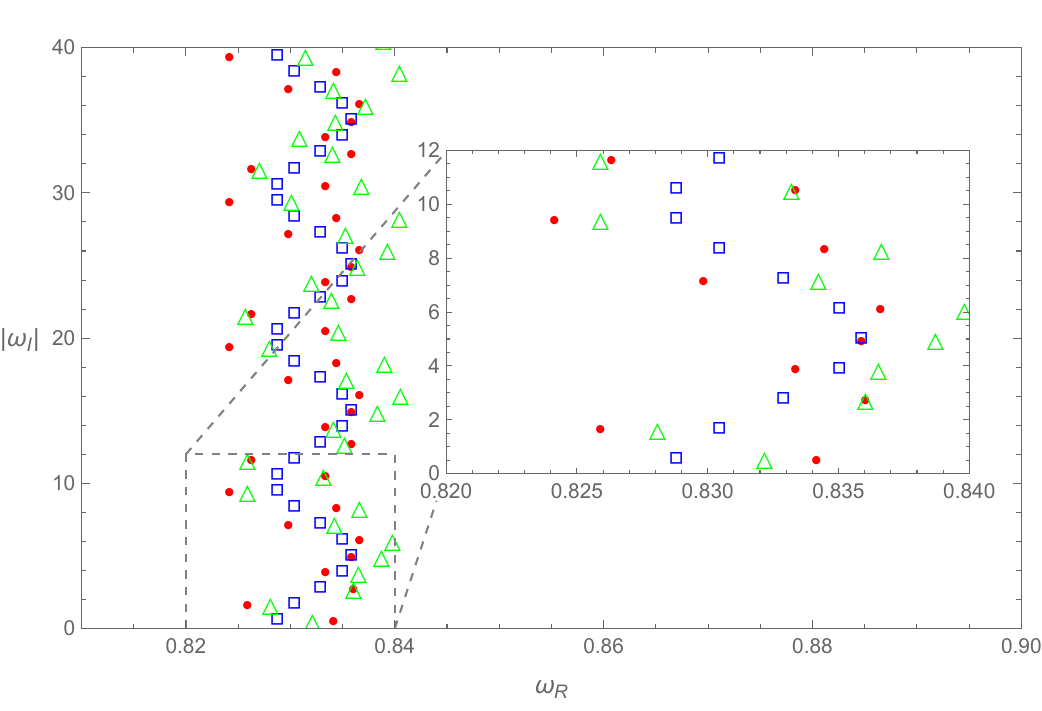}
\caption{The same as Fig.~\ref{fig1}, but the calculations are carried out using the metric parameters $x_c=0$, $(V_{0-}, b_{-}) = (1, 1)$, and $(V_{0+},b_{+})=(1,0.8)$.}
\label{fig3}
\end{figure}

Now, we are in a position to present the numerical results.
For different metric parameters, in Figs.~\ref{fig1}-\ref{fig3}, we present the asymptotic QNMs evaluated using the matrix method and compared with those obtained in~\cite{agr-qnm-Poschl-Teller-03, agr-qnm-Poschl-Teller-04}.
The numerical results obtained by the matrix method are shown in filled red circles, and the semi-analytic results given in~\cite{agr-qnm-Poschl-Teller-03, agr-qnm-Poschl-Teller-04} are represented by empty blue squares.
The matrix method is implemented using different grid sizes $N=90$, $120$, and $150$.
Subsequently, superfluous modes pertaining to the specific grid choice are removed by selecting only the modes that coincide.
Our results, particularly the detailed deviation between the two approaches, are further scrutinized using a semi-analytic approach of higher order elaborated later, indicated by empty green triangles.

Observing Figs.~\ref{fig1}-\ref{fig3}, it is apparent that all three approaches give primarily consistent results.
This is particularly the case when the size of the discontinuity, measured by the difference between $b_+$ and $b_-$, is insignificant.
However, the deviation becomes more significant when the discontinuity becomes more pronounced.
Specifically, as one compares the size of the deviation among the three figures, sizable deviations are observed in Fig.~\ref{fig3} when compared with those presented in Figs.~\ref{fig1} and~\ref{fig2}.

Besides, the derivations given in~\cite{agr-qnm-Poschl-Teller-03, agr-qnm-Poschl-Teller-04} are based on the assumption that the quasinormal frequencies have a significant imaginary part, namely, $|\Im\omega_n| \gg\Re\omega_n$.
This implies that the approximation might become less accurate for the low-lying modes.
This is indeed manifested by the figures' insets, as the deviation between the numerical and semi-analytic results is more significant for the low-lying modes closer to the real axis.

Moreover, we notice the deviation that the deviation occurs mainly for the real part of the frequencies, while their signs flip between the even and odd modes in accordance with the parity of the corresponding waveforms (see the example of two wave functions with even and odd $n$ in Appx.~\ref{appA}).
This parity-dependent deviation will be further elaborated in Sec.~\ref{sec5} and Appx.~\ref{appA}.
However, we note that such a feature is not indicated by Eq.~\eqref{Visser_result1}.
As mentioned above, the second line of Eq.~\eqref{Visser_result1} can be used as an iterative relation if one views the superscript $(1)$ and $(0)$ as $(j)$th and $(j-1)$th order.
Specifically, one can repeatedly substitute the obtained quasinormal frequency for the expression's r.h.s. to find a higher-order result.
Numerically, however, the frequencies evaluated according to Eq.~\eqref{Visser_result1} converge rather rapidly.
Based on the numerical values given in Tab.~\ref{Tab.1}, the obtained frequencies with ten iterations are essentially indistinguishable from their first-order counterparts.
The observed discrepancies in Figs.~\ref{fig1}-\ref{fig3} and their reason will be further explored in Sec.~\ref{sec5}, where we performed a refined version of the approximation that properly accounts for the observed deviations.

\begin{table}[htbp] 
\caption{\label{Tab.1} The resulting values of QNMs obtained using the relation Eq.~\eqref{Visser_result1} for different numbers of iterations.} 
\adjustbox{max width=\textwidth}{%
\begin{tabular}{|c|ccc|}
\hline
      $n$ &$\omega_{n}^{(5)}$&$\omega_{n}^{(10)}$&$\omega_{n}^{(15)}$  \\
\hline
     $0$&$~0.864565571423-0.502511946325i~$&$~0.864565571423-0.502511946325i~$&$~0.864565571423-0.502511946325i~$  \\
     $1$&$~0.864567004829-1.507535839602i~$&$~0.864567004829-1.507535839602i~$&$~0.864567004829-1.507535839602i~$  \\
     $2$&$~0.864569870172-2.512559734763i~$&$~0.864569870172-2.512559734763i~$&$~0.864569870172-2.512559734763i~$  \\
     $3$&$~0.864574164520-3.517583633061i~$&$~0.864574164520-3.517583633061i~$&$~0.864574164520-3.517583633061i~$  \\
     $4$&$~0.864579883477-4.522607535746i~$&$~0.864579883477-4.522607535746i~$&$~0.864579883477-4.522607535746i~$  \\
     $5$&$~0.864587021190-5.527631444062i~$&$~0.864587021190-5.527631444062i~$&$~0.864587021190-5.527631444062i~$  \\
     $6$&$~0.864595570355-6.532655359249i~$&$~0.864595570355-6.532655359249i~$&$~0.864595570355-6.532655359249i~$  \\
     $7$&$~0.864605522223-7.537679282539i~$&$~0.864605522223-7.537679282539i~$&$~0.864605522223-7.537679282539i~$  \\
     $8$&$~0.864616866614-8.542703215154i~$&$~0.864616866614-8.542703215154i~$&$~0.864616866614-8.542703215154i~$  \\
     $9$&$~0.864629591925-9.547727158308i~$&$~0.864629591925-9.547727158308i~$&$~0.864629591925-9.547727158308i~$  \\
     $10$&$~0.864643685141-10.55275111320i~$&$~0.864643685141-10.55275111320i~$&$~0.864643685141-10.55275111320i~$  \\
\hline
\end{tabular}}
\end{table}

\begin{figure}[htp]
\centering
\includegraphics[scale=0.65]{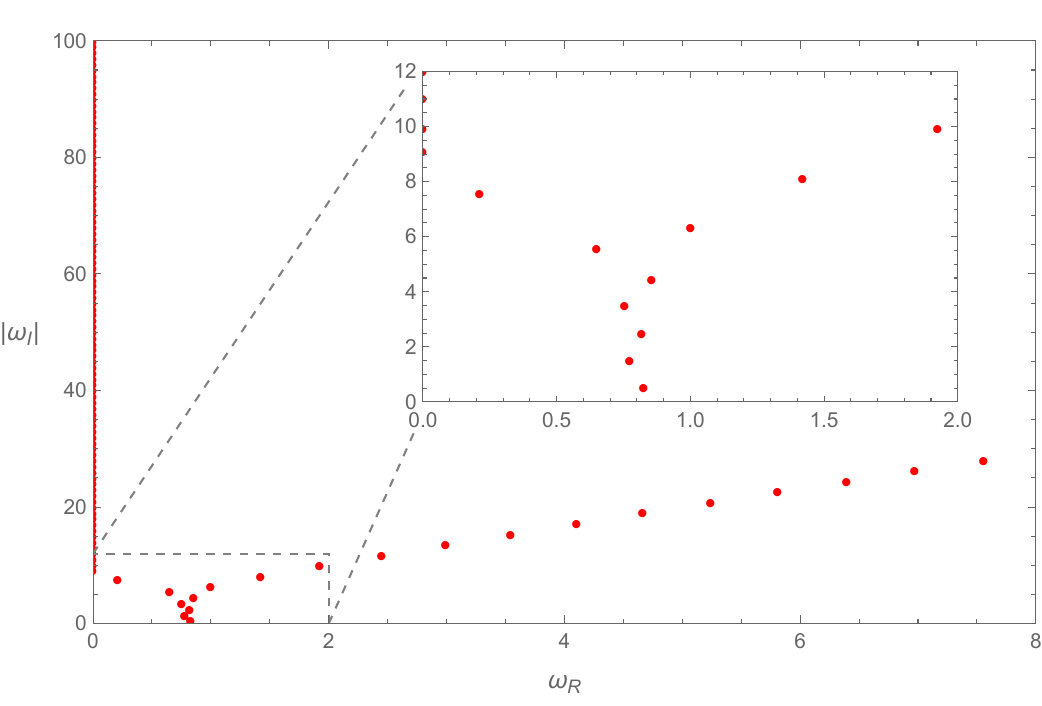}
\caption{Asymptotic QNMs for the modified P\"oschl-Teller effective potential Eq.~\eqref{Veff_MPT}. 
The calculations are carried out using the metric parameters $x_c=0.5$, $(V_{0-}, b_{-})=(1, 1)$, and $(V_{0+}, b_{+}) = (1-0.5^2, 1)$.
This corresponds to the scenario explored in~\cite{agr-qnm-lq-03} where an insignificant {\it step} is placed away from the origin.
The numerical results obtained by the matrix method discussed in Sec.~\ref{sec3} are shown in filled red circles, and the obtained quasinormal frequencies are verified by numerically evaluating the Wronskian of the waveforms, as discussed in the text.
The inset of the figure shows a zoomed-in section of a few lowest-lying modes indicated by the dashed square box.}
\label{fig4}
\end{figure}

\begin{figure}[htp]
\centering
\includegraphics[scale=0.65]{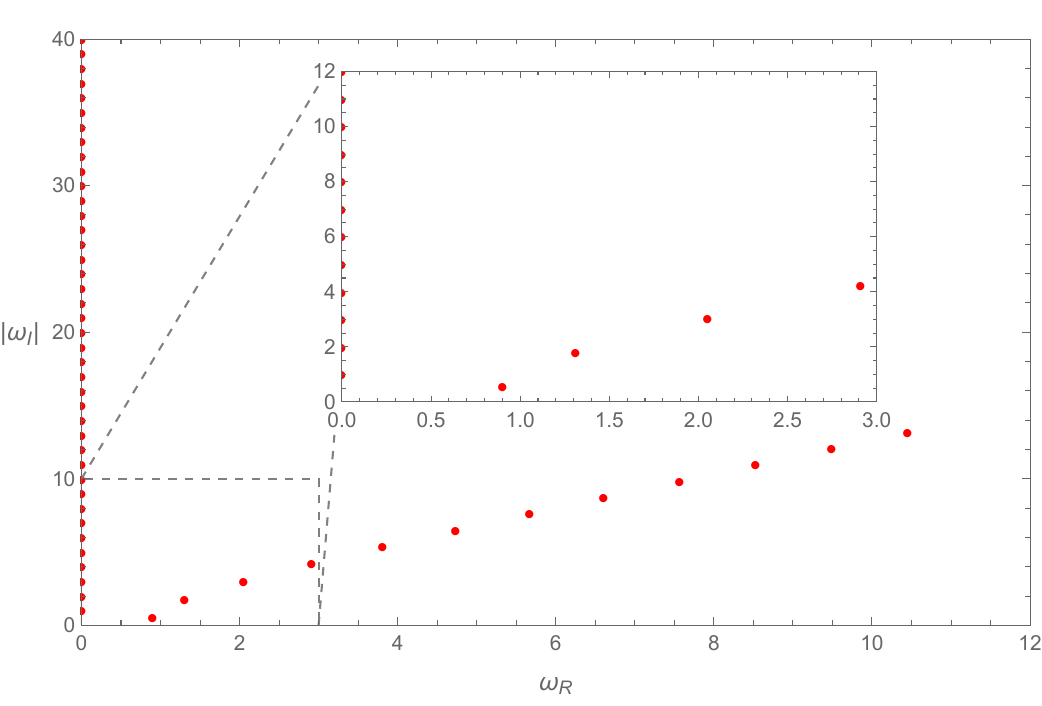}
\caption{The same as Fig.~\ref{fig1}, but the calculations are carried out using the metric parameters $(V_{0-}, b_{-}) = (1, 1)$, and $(V_{0+}, b_{+}) = (0, 1)$.
This corresponds to the scenario explored in~\cite{agr-qnm-lq-03} where the effective potential is {\it truncated} at a point further away from the origin.}
\label{fig5}
\end{figure}

We proceed to discuss the scenarios when a minor discontinuity is placed away from the origin.
In Figs.~\ref{fig4} and~\ref{fig5}, we first present two typical scenarios where a step and a truncation are introduced to the effective potential.
The matrix method is implemented using different grid sizes $N=90$, $120$, and $150$.
Again, unphysical modes are removed by retaining the ones that persist for different grids.

For both cases, one observes a branch of asymptotic modes extending along the real axis, which can be understood in terms of spectral instability~\cite{agr-qnm-instability-07}.
Moreover, there is another branch of purely imaginary modes sitting on the vertical axis.
In Fig.~\ref{fig4}, it is apparent that the two branches originate from a bifurcation point, below which the modes of the original P\"oschl-Teller potential remain largely intact.
The asymptotic modes along the real axis are not a novelty~\cite{agr-qnm-35, agr-qnm-36, agr-qnm-50} and can be readily understood by explicit semi-analytic calculations~\cite{agr-qnm-lq-03}.
However, to our knowledge, the purely imaginary modes have yet to be reported in the literature. 
To ascertain the numerical results obtained by the matrix method, we verify the quasinormal modes by explicitly evaluating the waveforms, which is further elaborated in Appx.~\ref{appA}.
For the case of the step, as shown in Fig.~\ref{fig4}, the QNMs correspond to the roots of the Wronskian Eq.~\eqref{WronskianForm} between the ingoing and outgoing waveforms.
However, for the truncated effective potential shown in Fig.~\ref{fig5}, the Wronskian diverges at the position of QNMs.
For this particular scenario, one must employ the junction condition given by Eq.~\eqref{junctionCondition}.

\begin{figure}[htp]
\centering
\includegraphics[scale=0.5]{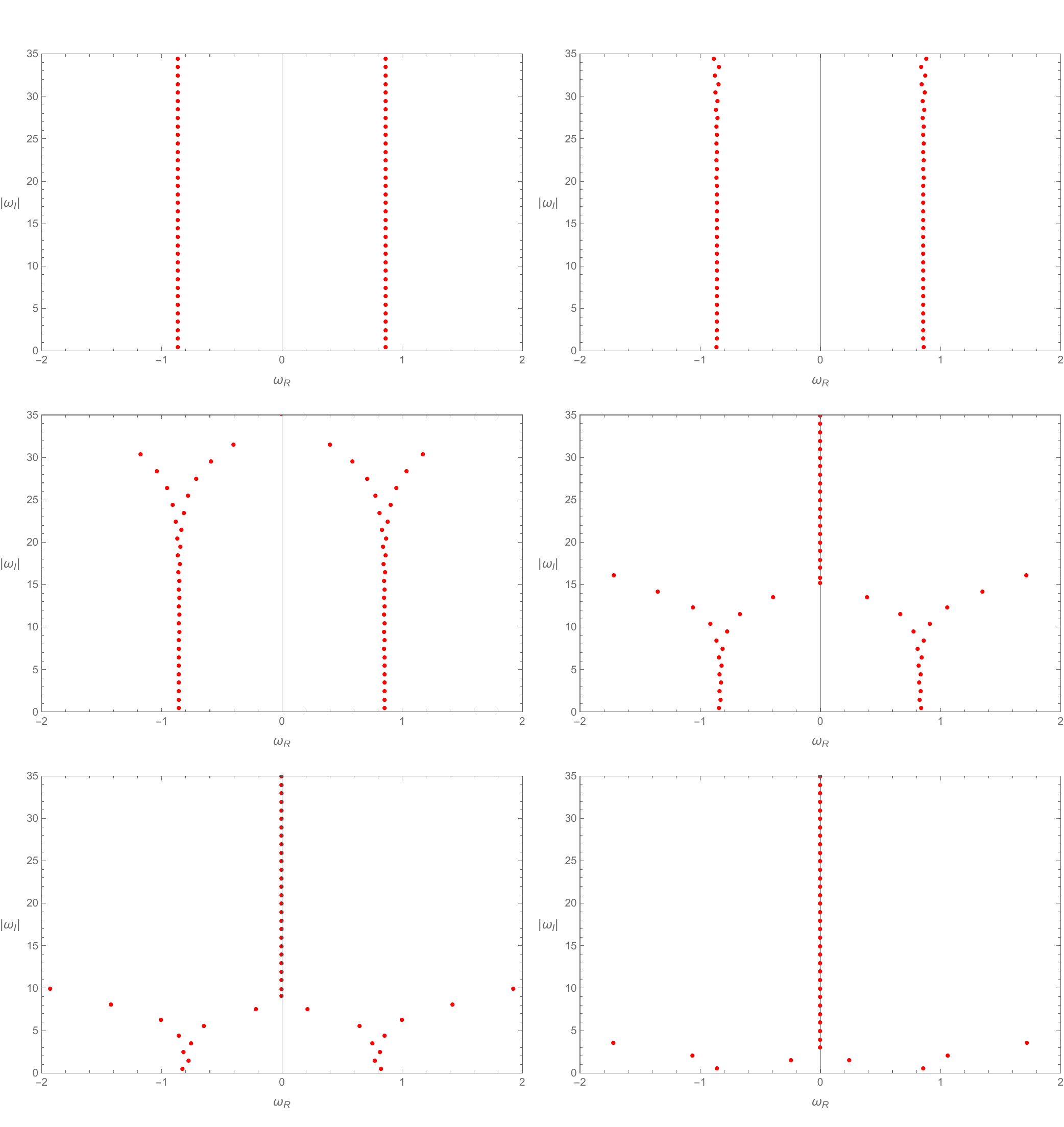}
\caption{Evolution of the QNMs for the modified P\"oschl-Teller effective potential Eq.~\eqref{Veff_MPT}. 
The calculations are carried out using the metric parameters $(V_{0-}, b_{-})=(1, 1)$ and $b_{+} = 1$ while varying $x_c$ and subsequently $V_{0+}$ by Eq.~\eqref{x0vary}.
This corresponds to a gradual evolution from the scenario explored in~\cite{agr-qnm-Poschl-Teller-03, agr-qnm-Poschl-Teller-04} to that discussed in ~\cite{agr-qnm-lq-03} when an insignificant discontinuity is moving away from the black hole.
The discontinuity is placed at $x_c = 0, 0.15, 0.2, 0.35, 0.5$, and $0.8$ from the top left to the bottom right panels.
The numerical results obtained by the matrix method discussed in Sec.~\ref{sec3} are shown in filled red circles, and the obtained quasinormal frequencies are verified by numerically evaluating the Wronskian of the waveforms, as discussed in the text.}
\label{fig6}
\end{figure}

It is apparent that the results shown in Figs.~\ref{fig1}-\ref{fig3} and Figs.~\ref{fig4}-\ref{fig5} are drastically different.
Therefore, it leads to the question of how they might be related.
This aspect can be explored by gradually evolving the location of a discontinuity from the origin to the spatial infinity while the height of the potential is suppressed during the process by the following relation
\bqn
V_{0+} = V_{0-}(1-x_c^2) .
\lb{x0vary}
\eqn
The resulting QNM spectra are presented in Fig.~\ref{fig6}.
It presents a gradual evolution of QNMs for the modified P\"oschl-Teller effective potential Eq.~\eqref{Veff_MPT} from the scenario elaborated in~\cite{agr-qnm-Poschl-Teller-03, agr-qnm-Poschl-Teller-04} to that discussed in ~\cite{agr-qnm-lq-03} when an insignificant discontinuity is moving away from the black hole.
Fig.~\ref{fig6} shows that the bifurcation point emerges from the high overtones, where the system is more susceptible to spectral instability.
This feature is somewhat reminiscent of the results found for randomized metric perturbations~\cite{agr-qnm-instability-19}.
However, rather than moving away from the imaginary axis and falling onto the real axis, a bifurcation occurs.
The plots show that the triggered instability propagates from high overtones to low-lying modes.
The modes falling towards the imaginary axis merge into the ones lying precisely on the imaginary axis.

Before closing this section, we note that the modes on the imaginary axis can be obtained using a semi-analytic approach.
Detailed derivations, inclusively the locations of these modes, will be relegated to Sec.~\ref{sec6}.

\section{A refined semi-analytic approach for the asymptotic quasinormal modes in the modified P\"oschl-Teller potential}\label{sec5}

In this section, we generalize the derivations of the asymptotic QNMs in the modified P\"oschl-Teller potential to the second order.
As discussed above in Sec.~\ref{sec4} and shown in Figs.~\ref{fig1}-\ref{fig3}, these corrections reasonably account for the observed deviations found between semi-analytic results Eq.~\eqref{Visser_result1} and those obtained by the matrix method.

For the QNMs of the P\"oschl-Teller potential in the hyperboloidal coordinates, one substitutes $x=1-2z$ into Eq.~\eqref{simplified_master_equation_PT} and finds 
\bqn
\left[z(1-z)\partial_{z,z}+(2z-1)(ib\omega-1)\partial_{z}+b(b(\omega^2-V_0)+i\omega)\right]\psi=0 ,
\lb{simplified_master_equation_PT_x=1-2z}
\eqn
which is Euler's hypergeometric differential equation~\cite{book-methods-mathematical-physics-09}.
\bqn
\left[z(1-z)\partial_{z,z}+(C-(A+B+1)z)\partial_{z}-AB\right]\psi=0,
\lb{Euler_hypergeometric_differential_equation}
\eqn
where
\bqn
A=-ib\omega+\frac{1}{2}\pm\alpha,~~~~B=-ib\omega+\frac{1}{2}\mp\alpha,&~&~~~C=-ib\omega+1 .
\lb{Euler_equation_solution}
\eqn

The solution is the hypergeometric function, which can be expressed by the power series
\bqn
\psi(\omega, x)={_{2}F_1}(A,B,C,z)=\sum_{n=0}^{+\infty} \beta_nz^n,
\lb{solution_before_trunction}
\eqn
where the coefficients of the hypergeometric series satisfy the iterative relation
\bqn
(n+1)(C+1)\beta_{n+1}&=&(A+n)(B+n)\beta_n .\lb{CorrHypergeometricSeries}
\eqn
Now, the QNM problem is different from the standard scenario when the wave functions are expressed in the spatial coordinates of an asymptotically flat spacetime. 
As discussed in Appx.~\ref{appA}, in such a case, the waveforms are subjected to the outgoing wave boundary condition Eq.~\eqref{master_bc0}. 
In the hyperboloidal coordinates, however, one only demands that the wave function be finite at the boundaries~\cite{agr-qnm-hyperboloidal-01} $z\to 0$ and $1$ where the hypergeometric function is typically divergent. 
Finiteness can be ensured if the hypergeometric series is truncated at the finite order.
By observing Eq.~\eqref{CorrHypergeometricSeries}, a truncation occurs for $A=-n$ or $B=-n$. 
Setting $A=-n$, one finds the condition
\bqn
-ib\omega+\frac{1}{2}\pm\alpha=-n,~~~~-ib\omega+\frac{1}{2}\mp\alpha=-2ib\omega+n+1 ,\lb{truncated_relations}
\eqn
which readily gives the QNM frequencies of the P\"ochl-Teller potential $\omega_n^\mathrm{PT}$ given by Eq.~\eqref{qnm_PT}.
By setting $B=-n$, identical results are retrieved.
As discussed in Appx.~\ref{appA}, the resulting waveforms possess well-defined parity $(-1)^n$.

Now, we consider the effective potential Eq.~\eqref{Veff_MPT} with $x_c=0$, $V_{0-}=V_{0+}$, and $b_{-}\simeq b_{+}$.
The waveforms on both sides of the discontinuity read from Eq.~\eqref{solution_before_trunction}
\bqn
\psi_{\pm}(\omega, x)={_{2}F_1}\left(-ib_{\pm}\omega+\frac{1}{2}\pm\alpha_{\pm},-ib_{\pm}\omega+\frac{1}{2}\mp\alpha_{\pm},-ib_{\pm}\omega+1,\frac{1\mp x}{2}\right),
\lb{discontinuous_PT_wavefunction}
\eqn
and their derivatives are found to be
\bqn
\partial_{x}\psi_{\pm}(\omega, x)=\mp\frac{\left(-ib_{\pm}\omega+\frac{1}{2}\right)^2-\alpha_{\pm}^2}{2(-ib_{\pm}\omega+1)}{_{2}F_1}\left(-ib_{\pm}\omega+\frac{3}{2}\pm\alpha_{\pm},-ib\omega_{\pm}+\frac{3}{2}\mp\alpha_{\pm},-ib_{\pm}\omega+2,\frac{1\mp x}{2}\right),\nb\\
\lb{differential_discontinuous_PT_wavefunction}
\eqn

The junction condition Eq.~\eqref{junctionCondition} at $x_c=0$ gives
\bqn
\left.\frac{\partial_{x}\psi_{-}(\omega, x)}{\psi_{-}(\omega, x)}\right|_{x=0}=\left.\frac{\partial_{x}\psi_{+}(\omega, x)}{\psi_{+}(\omega, x)}\right|_{x=0}.
\lb{boundary_condition_x=0}
\eqn
As shown in Appx.~\ref{appB}, it can be simplified to the following form
\bqn
{\cos}\left(-2\pi i \omega b_*\right)={\cos}\left(-\pi i \omega\Delta\right)-2{\cos(\pi\alpha_{-})}{\cos(\pi\alpha_{+})}+T_\mathrm{corr}(b_{\pm},\alpha_{\pm},\omega),\nb\\
\lb{high_order_result}
\eqn
where $\Delta$ and $b_*$ are defined by Eq.~\eqref{Visser_parameters2}, and the correction $T_\mathrm{corr}(b_{\pm},\alpha_{\pm},\omega)$ is given by
\bqn
T_\mathrm{corr}(b_{\pm},\alpha_{\pm},\omega)=2\frac{\sqrt{\left(ib_{-}\omega-\frac{1}{2}\right)^2-\alpha_{-}^2}-\sqrt{\left(ib_{+}\omega-\frac{1}{2}\right)^2-\alpha_{+}^2}}{\sqrt{\left(ib_{-}\omega-\frac{1}{2}\right)^2-\alpha_{-}^2}+\sqrt{\left(ib_{+}\omega-\frac{1}{2}\right)^2-\alpha_{+}^2}}\left(\cos(\pi\alpha_{-})\sin(\pi ib_{+}\omega)-\cos(\pi\alpha_{+})\sin(\pi ib_{-}\omega)\right).\nb\\
\lb{correction_term}
\eqn
In the derivation, specific properties of the hypergeometric and Gamma functions are utilized.
Specifically, one asymptotically expands the Gamma function in a region where it is well-behaved by making use of the fact that one is interested in asymptotic modes $\Im\omega \gg \Re\omega$, closely following the techniques developed in~\cite{agr-qnm-Poschl-Teller-03, agr-qnm-Poschl-Teller-04}.

Eq.~\eqref{high_order_result} is the desired approximate equation whose roots govern the asymptotic QNMs.
The frequency has a significant imaginary part $|\Im\omega| \gg 1$, so the resultant modulation is most pronounced for trigonometric functions.
Moreover, as pointed out in~\cite{agr-qnm-Poschl-Teller-03}, since $\Delta$ is insignificant, a sinusoidal function with the argument $\omega\Delta$ plays a less significant role compared to that with $\omega b_*$.
In other words, the r.h.s. of Eq.~\eqref{high_order_result} is a relatively moderate function in $ \omega $ concerning the l.h.s. of the equation.
When compared against the existing result Eq.~\eqref{Visser_QNF_condition}, the term $T_\mathrm{corr}(b_{\pm},\alpha_{\pm},\omega)$ corresponds to a sizable correction which is parity dependent.
To prove our claim, let us substitute the zeroth order result, Eq.~\eqref{Visser_result0}, into the r.h.s. of Eq.~\eqref{high_order_result}.
We have
\bqn
\omega_n^{(1)}=-i\frac{\arccos\left(\cos(-i\pi{\omega}_n^{(0)}\Delta)-2{\cos(\pi\alpha_{-})}{\cos(\pi\alpha_{+})}+ T_\mathrm{corr} (b_{\pm},\alpha_{\pm},{\omega}_n^{(0)})\right)}{2\pi b_{*}}-i\frac{n}{b_{*}}.
\lb{Novo_first_order_result}
\eqn
The correction $T_\mathrm{corr} (b_{\pm},\alpha_{\pm},{\omega}_n^{(0)})$ is manifestly insignificant because 
\bqn
\sin(\pi i b_\pm\omega_n^{(0)})
=\sin\left(\pi i \left(b_*\mp\frac{\Delta}{2}\right)\omega_n^{(0)}\right)
\simeq \sin\left(n\pi\mp\frac{n\pi\Delta}{2b_*} \right)
=\mp(-1)^n\sin\left(\frac{n\pi\Delta}{2b_*} \right)
\simeq\mp(-1)^n\left(\frac{n\pi\Delta}{2b_*} \right) .\nb\\
\lb{QNMcorrection}
\eqn
Interestingly, one observes that the factor $(-1)^n$ dictates that such a correction to the quasinormal frequencies Eq.~\eqref{Novo_first_order_result} is parity dependent.
Moreover, although the correction is the first-order one proportional to $\Delta$,
it overwhelms that due to the term $\cos(-i\pi{\omega}_n^{(0)}\Delta)$, which effectively is of second-order as a rectification to the argument of the overall $\arccos()$ operation.
These properties are indicated in Figs.~\ref{fig1}-\ref{fig3}.
We also note that Eq.~\eqref{Novo_first_order_result} can be viewed as an iterative relation. 
The convergence of~Eq.~\eqref{Novo_first_order_result} is manifestly fast, as seen in Tab.~\eqref{Tab.2}. 
The iterative results of Eq.~\eqref{Visser_result1} are shown in Tab.~\eqref{Tab.1}, and its convergence speed is relatively fast, but it can be seen that it cannot produce the results of two branches after iteration.

\begin{table}[htbp] 
\caption{\label{Tab.2} The resulting values of QNMs obtained using the relation Eq.~\eqref{Novo_first_order_result} for different numbers of iterations.} 
\adjustbox{max width=\textwidth}{%
\begin{tabular}{|c|ccc|}
\hline
    $n$  &$\omega_{n}^{(5)}$&$\omega_{n}^{(10)}$&$\omega_{n}^{(15)}$  \\
\hline
     $0$&$~0.864571102938-0.502502429535i~$&$~0.864571102938-0.502502429535i~$&$~0.864571102938-0.502502429535i~$  \\
     $1$&$~0.864561999703-1.507507369767i~$&$~0.864561999703-1.507507369767i~$&$~0.864561999703-1.507507369767i~$  \\
     $2$&$~0.864574797045-2.512504067014i~$&$~0.864574797045-2.512504067014i~$&$~0.864574797045-2.512504067014i~$  \\
     $3$&$~0.864569098878-3.517512402062i~$&$~0.864569098878-3.517512402062i~$&$~0.864569098878-3.517512402062i~$  \\
     $4$&$~0.864585205342-4.522503129987i~$&$~0.864585205342-4.522503129987i~$&$~0.864585205342-4.522503129987i~$  \\
     $5$&$~0.864581594260-5.527515750115i~$&$~0.864581594260-5.527515750115i~$&$~0.864581594260-5.527515750115i~$  \\
     $6$&$~0.864601265237-6.532501286901i~$&$~0.864601265237-6.532501286901i~$&$~0.864601265237-6.532501286901i~$  \\
     $7$&$~0.864599871566-7.537518580349i~$&$~0.864599871566-7.537518580349i~$&$~0.864599871566-7.537518580349i~$  \\
     $8$&$~0.864622819936-8.542499243060i~$&$~0.864622819936-8.542499243060i~$&$~0.864622819936-8.542499243060i~$  \\
     $9$&$~0.864623848084-9.547521300507i~$&$~0.864623848084-9.547521300507i~$&$~0.864623848084-9.547521300507i~$  \\
     $10$&$~0.864649811709-10.55249727655i~$&$~0.864649811709-10.55249727655i~$&$~0.864649811709-10.55249727655i~$  \\
\hline
\end{tabular}}
\end{table}

\section{A semi-analytic derivation for purely imaginary quasinormal modes for truncated P\"oschl-Teller potential}\label{sec6}

This section discusses the purely imaginary modes on the imaginary axis found numerically in Sec.~\ref{sec4}.
Although a general analytic proof of the location of the pure imaginary modes does not seem straightforward, it is possible to formulate a proof for asymptotic modes.

We consider that the truncation is located at $x_c$ and expands the waveforms around the origin, namely, $x_c=x_0+\Delta x$ for $x_0=0$.
While delegating a general proof to Appx.~\ref{appB}, here we give an account of an approximate but explicit approach.
At $x_c$, the junction condition~\eqref{junctionCondition} reads
\bqn
\left.\frac{\partial_{x}\psi_{-}(\omega,x)}{\psi_{-}(\omega,x)}\right|_{x_c=x_0+\Delta x}=\left.\frac{\partial_{x}\psi_{+}(\omega,x)}{\psi_{+}(\omega,x)}\right|_{x_c=x_0+\Delta x} .
\lb{boundary_condition_x0=x1-Deltax}
\eqn
In the region where the effective potential is truncated $x > x_c$, we consider the following metric parameters 
\bqn
(V_{0+}, b_{+}) = (0, b_{-}), \lb{truncPar}
\eqn

One explicitly expands both sides of Eq.~\eqref{boundary_condition_x0=x1-Deltax} to the third order, which suffices if the truncation point is not far from the origin.
We have
\bqn
&&\left.\frac{\partial_{x}\psi_{\pm}(\omega,x)}{\psi_{\pm}(\omega,x)}\right|_{x_c}=\mp2\frac{\Gamma\left(\frac{1}{2}(1-A_{\pm})+\frac{1}{2}\right)}{\Gamma\left(\frac{1}{2}(1-A_{\pm})\right)}\frac{\Gamma\left(\frac{1}{2}(1-B_{\pm})+\frac{1}{2}\right)}{\Gamma\left(\frac{1}{2}(1-B_{\pm})\right)}\tan(\frac{\pi}{2}A_{\pm})\tan(\frac{\pi}{2}B_{\pm})\nonumber\\
&&+\left((A_{\pm})(B_{\pm})-\left(\mp2\frac{\Gamma\left(\frac{1}{2}(1-A_{\pm})+\frac{1}{2}\right)}{\Gamma\left(\frac{1}{2}(1-A_{\pm})\right)}\frac{\Gamma\left(\frac{1}{2}(1-B_{\pm})+\frac{1}{2}\right)}{\Gamma\left(\frac{1}{2}(1-B_{\pm})\right)}\tan(\frac{\pi}{2}A_{\pm})\tan(\frac{\pi}{2}B_{\pm})\right)^2\right)\Delta x\nonumber\\
&&+\frac{1}{2}\Bigg(\mp2\big((A_{\pm}+1)(B_{\pm}+1)-(A_{\pm})(B_{\pm})\big)\frac{\Gamma\left(\frac{1}{2}(1-A_{\pm})+\frac{1}{2}\right)}{\Gamma\left(\frac{1}{2}(1-A_{\pm})\right)}\frac{\Gamma\left(\frac{1}{2}(1-B_{\pm})+\frac{1}{2}\right)}{\Gamma\left(\frac{1}{2}(1-B_{\pm})\right)}\tan(\frac{\pi}{2}A_{\pm})\tan(\frac{\pi}{2}B_{\pm})\nonumber\\
&&~~~~~~~-2\Bigg(\mp2(A_{\pm})(B_{\pm})\frac{\Gamma\left(\frac{1}{2}(1-A_{\pm})+\frac{1}{2}\right)}{\Gamma\left(\frac{1}{2}(1-A_{\pm})\right)}\frac{\Gamma\left(\frac{1}{2}(1-B_{\pm})+\frac{1}{2}\right)}{\Gamma\left(\frac{1}{2}(1-B_{\pm})\right)}\tan(\frac{\pi}{2}A_{\pm})\tan(\frac{\pi}{2}B_{\pm})\nonumber\\
&&~~~~~~~-\left(\mp2\frac{\Gamma\left(\frac{1}{2}(1-A_{\pm})+\frac{1}{2}\right)}{\Gamma\left(\frac{1}{2}(1-A_{\pm})\right)}\frac{\Gamma\left(\frac{1}{2}(1-B_{\pm})+\frac{1}{2}\right)}{\Gamma\left(\frac{1}{2}(1-B_{\pm})\right)}\tan(\frac{\pi}{2}A_{\pm})\tan(\frac{\pi}{2}B_{\pm})\right)^3\Bigg)(\Delta x)^2,
\lb{dpsi/psi_x1=0_equation}
\eqn
where $A_{\pm},B_{\pm},C_{\pm}$ are defined by Eq.~\eqref{Euler_equation_solution}.
To proceed, one substitutes the specific metric parameters for the waveforms on both sides of the truncation.
Besides, as in the last section, one employs Eq.~\eqref{avoid_poles} to avoid the region in the complex plane where the Gamma function possesses many poles.
For asymptotic modes whose imaginary part is negative and significant, one makes use of Eq.~\eqref{Gam_approximate} to simplify the above equation into 
\bqn
&&\left.\frac{\partial_{x}\psi_{\pm}(\omega,x)}{\psi_{\pm}(\omega,x)}\right|_{x_c}\simeq\mp\sqrt{(1-A_{\pm})(1-B_{\pm})}\tan\left(\frac{\pi}{2}A_{\pm}\right)\tan\left(\frac{\pi}{2}B_{\pm}\right)\nonumber\\
&&~~~~~~~~~~~~~~~~~~~~~~+\left((A_{\pm})(B_{\pm})-(1-A_{\pm})(1-B_{\pm})\left(\tan\left(\frac{\pi}{2}A_{\pm}\right)\tan\left(\frac{\pi}{2}B_{\pm}\right)\right)^2\right)\Delta x\nonumber\\
&&~~~~~~~~~~~~~~~~~~~~~~+\frac{1}{2}\Bigg(\mp\big((A_{\pm}+1)(B_{\pm}+1)-(A_{\pm})(B_{\pm})\big)\sqrt{(1-A_{\pm})(1-B_{\pm})}\tan\left(\frac{\pi}{2}A_{\pm}\right)\tan\left(\frac{\pi}{2}B_{\pm}\right)\nonumber\\
&&~~~~~~~~~~~~~~~~~~~~~~~~~~~~~~\pm2(A_{\pm})(B_{\pm})\sqrt{(1-A_{\pm})(1-B_{\pm})}\tan\left(\frac{\pi}{2}A_{\pm}\right)\tan\left(\frac{\pi}{2}B_{\pm}\right)\nonumber\\
&&~~~~~~~~~~~~~~~~~~~~~~~~~~~~~~\mp2\sqrt{(1-A_{\pm})^3(1-B_{\pm})^3}\left(\tan\left(\frac{\pi}{2}A_{\pm}\right)\tan\left(\frac{\pi}{2}B_{\pm}\right)\right)^3\Bigg)(\Delta x)^2.
\lb{dpsi/psi_x1=0_equation_1}
\eqn
For the l.h.s., it becomes
\bqn
&&\left.\frac{\partial_{x}\psi_{-}(\omega,x)}{\psi_{-}(\omega,x)}\right|_{x_c}=ib_{-}\omega\frac{{\cos}(\pi\alpha_{-})+{\sin}(-\pi ib_{-}\omega)}{{\cos}(\pi\alpha_{-})-{\sin}(-\pi ib_{-}\omega)}+\Bigg((ib_{-}\omega)^2\left(1-\left(\frac{{\cos}(\pi\alpha_{-})+{\sin}(-\pi ib_{-}\omega)}{{\cos}(\pi\alpha_{-})-{\sin}(-\pi ib_{-}\omega)}\right)^2\right)\nonumber\\
&&~~~~~~~~~~~~~~~~~~~~~~~-2ib_{-}\omega \left(\frac{{\cos}(\pi\alpha_{-})+{\sin}(-\pi ib_{-}\omega)}{{\cos}(\pi\alpha_{-})-{\sin}(-\pi ib_{-}\omega)}\right)^2\Bigg)\Delta x+\Bigg(-(ib_{-}\omega)^2\frac{{\cos}(\pi\alpha_{-})+{\sin}(-\pi ib_{-}\omega)}{{\cos}(\pi\alpha_{-})-{\sin}(-\pi ib_{-}\omega)}\nonumber\\
&&~~~~~~~~~~~~~~~~~~~~~~~-\bigg((ib_{-}\omega)^3\left(\frac{{\cos}(\pi\alpha_{-})+{\sin}(-\pi ib_{-}\omega)}{{\cos}(\pi\alpha_{-})-{\sin}(-\pi ib_{-}\omega)}-\left(\frac{{\cos}(\pi\alpha_{-})+{\sin}(-\pi ib_{-}\omega)}{{\cos}(\pi\alpha_{-})-{\sin}(-\pi ib_{-}\omega)}\right)^3\right)\nonumber\\
&&~~~~~~~~~~~~~~~~~~~~~~~-2(ib_{-}\omega)^2 \left(\frac{{\cos}(\pi\alpha_{-})+{\sin}(-\pi ib_{-}\omega)}{{\cos}(\pi\alpha_{-})-{\sin}(-\pi ib_{-}\omega)}\right)^3\bigg)\Bigg)(\Delta x)^2.
\lb{dpsi/psi_x1=0_equation_lhs}
\eqn
For the r.h.s., using Eq.~\eqref{truncPar}, we have
\bqn
&&\left.\frac{\partial_{x}\psi_{+}(\omega,x)}{\psi_{+}(\omega,x)}\right|_{x_c}=ib_{-}\omega-2ib_{-}\omega\Delta x+(ib_{-}\omega)^2(\Delta x)^2
\lb{dpsi/psi_x1=0_equation_rhs}
\eqn
Equating Eq.~\eqref{dpsi/psi_x1=0_equation_lhs} to~\eqref{dpsi/psi_x1=0_equation_rhs}, it is not difficult to observe that
\bqn
\omega=-i\frac{n}{b_{-}}. \lb{dpsi/psi_x0_final_2} ,
\eqn
are the roots of the junction condition since they imply
\bqn
\frac{{\cos}(\pi\alpha_{-})+{\sin}(-\pi ib_{-}\omega)}{{\cos}(\pi\alpha_{-})-{\sin}(-\pi ib_{-}\omega)}=1 ,
\lb{dpsi/psi_x0_final}
\eqn
which holds for any value of $\alpha_{-}$.

It is noted that when the truncation is placed at the origin, purely imaginary asymptotic modes were first derived in~\cite{agr-qnm-Poschl-Teller-03}. 
Our present findings are a further generalization.

\section{Concluding remarks}\label{sec7}

To summarize, this study aims to further explore properties of asymptotic QNMs in the modified P\"oschl-Teller potential under metric perturbations. 
While existing literature indicates a migration of QNMs toward the real axis in response to ultraviolet perturbations, there is seemingly ambiguity, as existing results also suggested that asymptotic modes may persist along the imaginary axis, particularly when the perturbation is placed at the origin.
By employing numerical and semi-analytical approaches, we have comprehensively analyzed the problem. 
Our numerical scheme, based on an improved matrix method implemented in hyperboloidal coordinates on the Chebyshev grid, confirms that, depending on the placement of the perturbation, both behaviors of asymptotic QNMs are observed. 
Notably, we report the emergence of a novel branch of purely imaginary modes originating from a bifurcation in the asymptotic QNM spectrum, shedding light on the dynamic emergence of spectral instability.
Moreover, our findings are reinforced by independent numerical verifications and semi-analytical approaches. 

The numerical calculations are primarily carried out using an improved version of the matrix method.
Compared to its predecessor, the present version can evaluate quasinormal frequencies of high overtones with desirable precision.
However, because the Wronkian involves frequency as the argument of the exponential function, the matrix method is shown to be even more capable of dealing with high overtones.
Meanwhile, the approach benefits from its effectiveness, revealing all possible QNMs simultaneously without demanding a high-precision initial trial value.
Also, for more general scenarios, it is understood that an analytic form of the Wronkian is not always accessible.
We note that some particular scenarios for purely imaginary modes have been discussed in~\cite{agr-qnm-Poschl-Teller-03}, consistent with our findings.
These modes have also been observed for randomized perturbations~\cite{agr-qnm-instability-07}.
Nonetheless, it is unclear whether the imaginary modes encountered in the present study can be expected in a more general context for perturbed metric or merely reflect a particular feature of the modified P\"oschl-Teller potential.
Besides, the physical content of such modes is obscure to us.
Further studies in this direction might be interesting.

Last but not least, the observational implications of our findings primarily reside on their potential impact on gravitational wave detection.
It has been speculated that the spectral instability might leave a detectable signature in the gravitational wave signals~\cite{agr-qnm-instability-13}.
This is understood as these asymptotic modes align themselves towards the real axis and might have a sizable collective effect~\cite{agr-qnm-echoes-20}. 
In this regard, it may imply further challenges for the ongoing endeavor of black hole spectroscopy, and in particular, explicit calculations for the signal-to-noise ratio~\cite{agr-TDI-review-02, agr-TDI-Wang-18} using arbitrary time-delay interferometry solutions~\cite{agr-TDI-Wang-09, agr-TDI-Wang-22} tailored for specific detector parameters might be pertinent.
We plan to continue exploring these topics in future studies.

\section*{Acknowledgements}

We extend our gratitude to Rui-Hong Yue, Kai Lin, and Stefan Randow for their insightful discussions.
We gratefully acknowledge the financial support from Brazilian agencies 
Funda\c{c}\~ao de Amparo \`a Pesquisa do Estado de S\~ao Paulo (FAPESP), 
Funda\c{c}\~ao de Amparo \`a Pesquisa do Estado do Rio de Janeiro (FAPERJ), 
Conselho Nacional de Desenvolvimento Cient\'{\i}fico e Tecnol\'ogico (CNPq), 
and Coordena\c{c}\~ao de Aperfei\c{c}oamento de Pessoal de N\'ivel Superior (CAPES).
This work is supported by the National Natural Science Foundation of China (NSFC).
GRL is supported by the China Scholarship Council.
A part of this work was developed under the project Institutos Nacionais de Ci\^{e}ncias e Tecnologia - F\'isica Nuclear e Aplica\c{c}\~{o}es (INCT/FNA) Proc. No. 464898/2014-5.
This research is also supported by the Center for Scientific Computing (NCC/GridUNESP) of S\~ao Paulo State University (UNESP).

\appendix

\section{The waveforms of the modified P\"oschl-Teller effective potential}\label{appA}

For the P\"oschl-Teller effective potential Eq.~\eqref{potential_PT}, one introduces two transforms
\bqn
\Psi(\omega, r_*) = \left(\frac{z}{1-z}\right)^{\frac{ib\omega}{2}} \varphi_\pm(\omega, z) ,
\eqn
where 
\bqn
z=\frac{1}{1+e^{\pm 2r_*/b}} .
\eqn
For both waveforms, the master equation Eq.~\eqref{master_frequency_domain} results in an identical form,
\bqn
z(1-z)\frac{d^2\varphi_\pm(\omega, z)}{dz^2}-(1 +i \omega b -2z) \frac{d\varphi_\pm(\omega, z)}{dz}- V_0 b^2\varphi_\pm(\omega, z) = 0 .
\eqn

It is readily recognized to be the hypergeometric differential equation Eq.~\eqref{Euler_hypergeometric_differential_equation} where
\bqn
A=\frac{1}{2}+\alpha,~~~~B=\frac{1}{2}-\alpha,&~&~~~C=1+ib\omega .
\lb{coefficients_Euler}
\eqn
There are two sets of independent solutions for $\Psi$ at our disposal, namely,
\bqn
\left(\frac{z}{1-z}\right)^{\frac{C-1}{2}}{_2}F_1\left(A, B, C, z\right) ,\nb
\eqn
and
\bqn
\left[(1-z)z\right]^{\frac{1-C}{2}}{_2}F_1\left(1+A-C, 1+B-C, 2-C, z\right) .\nb
\eqn
However, in asymptotical spacetime, choosing solutions that satisfy the outgoing boundary conditions Eq.~\eqref{master_bc0} at $z=\pm 1$ is convenient.
Noticing the specific value of the hypergeometric function
\bqn
\lim\limits_{z\to 0} {_2}F_1\left(A, B, C, z\right) = {_2}F_1\left(A, B, C, 0\right)= 1 ,\nb
\eqn
the two appropriate solutions are obtained by picking up one from each set, and we have~\cite{agr-qnm-Poschl-Teller-05}
\bqn
\Psi(\omega, r_*)_\pm = e^{\mp i\omega r_*} {_2}F_1\left(\frac{1}{2}+\alpha, \frac{1}{2}-\alpha, 1+ib\omega, \frac{1}{1+e^{\pm 2r_*/b}}\right) ,\lb{PsiRStar}
\eqn
where we note that $\left(\frac{z}{1-z}\right)^{\frac{C-1}{2}}=e^{\mp i\omega r_*}$.

\begin{figure}[htp]
\centering
\includegraphics[scale=0.5]{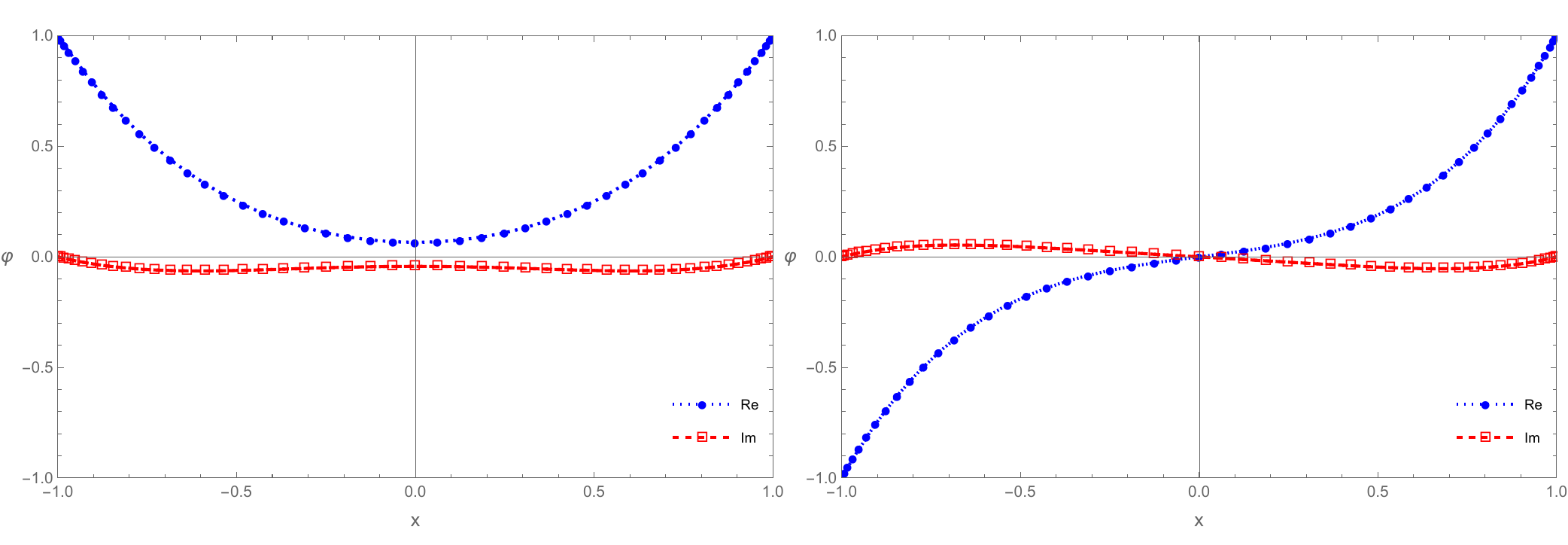}
\caption{The real and imaginary parts of QNM waveforms of the P\"oschl-Teller effective potential obtained by using the matrix method (shown in symbols), compared to the analytic result Eq.~\eqref{discontinuous_PT_wavefunction} (shown in dotted and dashed curves).
The calculations are carried out for the metric parameters $(b, V_{0})=(1,1)$ and presented in the hyperboloidal coordinates $x$.
The left panel corresponds to the even-parity waveform of the overtone $n=4$, and the right panel shows an odd-parity one with $n=5$.}
\label{fig7}
\end{figure}

Subsequently, for the quasinormal frequencies $\omega_n^\mathrm{PT}$ given by Eq.~\eqref{qnm_PT}, we have
\bqn
\Psi(\omega_n^\mathrm{PT}, r_*)=
\begin{cases}
   \Psi(\omega, r_*)_-, &  r_* < 0, \\
   (-1)^n\Psi(\omega, r_*)_+, &  r_* \ge 0,
\end{cases}
\lb{PT_wavefunctionRStar}
\eqn
It is worth pointing out that the waveforms Eqs.~\eqref{discontinuous_PT_wavefunction} and~\eqref{PT_wavefunctionRStar} all have a well-defined parity, as shown in Fig.~\ref{fig7}.
When a discontinuity is introduced to the potential, the deviations from these modes possess different signs for different parity, as shown numerically in Sec.~\ref{sec4} and semi-analytically in Sec.~\ref{sec5}. 

Generalizing the above result to the modified P\"oschl-Teller effective potential Eq.~\eqref{Veff_MPT} is straightforward.
Following very similar derivations, the resulting waveforms on both sides of the discontinuity read
\bqn
\Psi(\omega, r_*)_\pm = e^{\mp i\omega r_*} {_2}F_1\left(\frac{1}{2}+\alpha_\pm, \frac{1}{2}-\alpha_\pm, 1+ib_\pm\omega, \frac{1}{1+e^{\pm 2r_*/b_\pm}}\right) .
\eqn

By definition, the Wronskian involved in Sec.~\ref{sec4} reads
\bqn
W\left\{\Psi_-,\Psi_+\right\} = \left.\Psi_-{\Psi_+}'\right|_{x=x_c}- \left.\Psi_+{\Psi_-}'\right|_{x=x_c} ,\lb{WronskianForm}
\eqn
which must be evaluated at the discontinuity using the specific forms given above.

\section{Derivations for the asymptotic quasinormal modes for the modified P\"oschl-Teller effective potential}\label{appB}

This Appendix gives a detailed account of the analytic results presented in Secs.~\ref{sec5} and~\ref{sec6}.

We first derive the results for the perturbed modes in the modified P\"oschl-Teller effective potential utilized in Sec.~\ref{sec5}.
The derivative of the waveform 
\begin{equation}
\psi_{\pm}(\omega, x)={_{2}F_1}\left(-ib_{\pm}\omega+\frac{1}{2}\pm\alpha_{\pm},-ib_{\pm}\omega+\frac{1}{2}\mp\alpha_{\pm},-ib_{\pm}\omega+1,\frac{1\mp x}{2}\right), \tag{\ref{discontinuous_PT_wavefunction}}
\end{equation}
can be evaluated using the differentiation formula
\bqn
\frac{d^n}{dz^n}{_{2}F_1}\left(A, B, C, z\right)= \frac{(A)_n(B)_n}{(C)_n}{_{2}F_1}\left(A+n, B+n, C+n, z\right) ,\lb{diffForm}
\eqn
where $(\gamma)_n$ is the falling factorial is defined by
and the rising factorial is defined as 
\bqn
(H)_n=\prod_{k=0}^{n-1}(H+k) ,\lb{rFactorial}
\eqn
which gives 
\begin{equation}
\partial_{x}\psi_{\pm}(\omega, x)=\mp\frac{\left(-ib_{\pm}\omega+\frac{1}{2}\right)^2-\alpha_{\pm}^2}{2(-ib_{\pm}\omega+1)}{_{2}F_1}\left(-ib_{\pm}\omega+\frac{3}{2}\pm\alpha_{\pm},-ib\omega_{\pm}+\frac{3}{2}\mp\alpha_{\pm},-ib_{\pm}\omega+2,\frac{1\mp x}{2}\right) .  \tag{\ref{differential_discontinuous_PT_wavefunction}}
\end{equation}
Subsequently, their ratio reads
\bqn
\frac{\partial_{x}\psi_{-}(\omega, x)}{\psi_{-}(\omega, x)}=\mp\frac{\left(-ib_{\pm}\omega+\frac{1}{2}\right)^2-\alpha_{\pm}^2}{2(-ib_{\pm}\omega+1)}\frac{{_{2}F_1}\left(-ib_{\pm}\omega+\frac{3}{2}\pm\alpha_{\pm},-ib\omega_{\pm}+\frac{3}{2}\mp\alpha_{\pm},-ib_{\pm}\omega+2,\frac{1\mp x}{2}\right)}{{_{2}F_1}\left(-ib_{\pm}\omega+\frac{1}{2}\pm\alpha_{\pm},-ib_{\pm}\omega+\frac{1}{2}\mp\alpha_{\pm},-ib_{\pm}\omega+1,\frac{1\mp x}{2}\right)}.
\lb{dpsi/psi_x}
\eqn

Now we proceed to evaluate the junction condition Eq.~\eqref{junctionCondition} evaluated at $x_c=0$
\begin{equation}
\left.\frac{\partial_{x}\psi_{-}(\omega, x)}{\psi_{-}(\omega, x)}\right|_{x_c=0}=\left.\frac{\partial_{x}\psi_{+}(\omega, x)}{\psi_{+}(\omega, x)}\right|_{x_c=0}. \tag{\ref{boundary_condition_x=0}}
\end{equation}
We assume the waveforms and their derivatives do not vanish identically at the discontinuity.
By noticing the discontinuity $z=\frac12$, we make use of Gauss's second summation theorem:
\bqn
{_{2}F_1}\left(A,B,\frac{1}{2}(A+B+1),\frac{1}{2}\right)
=\frac{\Gamma(\frac{1}{2})\Gamma(\frac{1}{2}(A+B+1))}{\Gamma(\frac{1}{2}(A+1))\Gamma(\frac{1}{2}(B+1))}.
\lb{second_summation_theorem}
\eqn
As a result, both the waveforms and their derivatives can be rewritten as the product and ratios of a few Gamma functions.
Specifically, by substituting into Eqs.~\eqref{discontinuous_PT_wavefunction} and~\eqref{differential_discontinuous_PT_wavefunction} we find the exact result
\bqn
\frac{\partial_{x}\psi_{\pm}(\omega,0)}{\psi_{\pm}(\omega,0)}
=\mp\frac{\left(-ib_{\pm}\omega+\frac{1}{2}\right)^2-\alpha_{\pm}^2}{2(-ib_{\pm}\omega+1)}\frac{\Gamma\left(-ib_{\pm}\omega+2\right)}{\Gamma\left(-ib_{\pm}\omega+1\right)}\frac{\Gamma\left(\frac{1}{2}(-ib_{\pm}\omega+\frac{3}{2}\pm\alpha_{\pm})\right)}{\Gamma\left(\frac{1}{2}(-ib_{\pm}\omega+\frac{5}{2}\pm\alpha_{\pm})\right)}\frac{\Gamma\left(\frac{1}{2}(-ib_{\pm}\omega+\frac{3}{2}\mp\alpha_{\pm})\right)}{\Gamma\left(\frac{1}{2}(-ib_{\pm}\omega+\frac{5}{2}\mp\alpha_{\pm})\right)}.\nb\\
\lb{dpsi/psi_x=0_1}
\eqn

Since the asymptotic quasinormal modes possess a significant negative imaginary part satisfying $\left|\Im\omega_n\right|\to +\infty$, it implies that the Gamma function's argument sits at the region of the negative real axis where an asymptotic expansion is not feasible due to the presence of an array of poles.
Following~\cite{agr-qnm-Poschl-Teller-03, agr-qnm-Poschl-Teller-03}, one flips the argument to a region near the positive real axis where the Gamma function is well-behaved.
This is achieved by employing the Euler's reflection formula
\bqn
\Gamma(1-z)=\frac{\pi}{{\sin}(\pi z)\Gamma(z)},~~~~\mathrm{for}~~~~z\notin \mathbb{Z} ,
\lb{avoid_poles}
\eqn
through which the troublesome poles are carried by the trigonometric functions.
Utilizing the above strategy, one approximates Eq.~\eqref{dpsi/psi_x=0_1} to
\bqn
\frac{\partial_{x}\psi_{\pm}(\omega,0)}{\psi_{\pm}(\omega,0)}
\simeq\mp\sqrt{\left(ib_{\pm}\omega-\frac{1}{2}\right)^2-\alpha_{\pm}^2}\tan\left(\frac{\pi}{2}\left(-ib_{\pm}\omega+\frac{1}{2}\pm\alpha_{\pm}\right)\right)\tan\left(\frac{\pi}{2}\left(-ib_{\pm}\omega+\frac{1}{2}\mp\alpha_{\pm}\right)\right),
\lb{dpsi/psi_x=0_2}
\eqn
where, specifically, we have used
\bqn
&~~~&\frac{\Gamma\left(-ib_{\pm}\omega+2\right)}{\Gamma\left(-ib_{\pm}\omega+1\right)}=\frac{\Gamma\left(ib_{\pm}\omega\right)}{\Gamma\left(ib_{\pm}\omega-1\right)}\frac{{\sin}\left({\pi}(-ib_{\pm}\omega+1)\right)}{{\sin}\left({\pi}(-ib_{\pm}\omega+2)\right)}=-ib_{\pm}\omega+1,\nonumber\\
&~~~&\frac{\Gamma\left(\frac{1}{2}(-ib_{\pm}\omega+\frac{3}{2}\pm\alpha_{\pm})\right)}{\Gamma\left(\frac{1}{2}(-ib_{\pm}\omega+\frac{5}{2}\pm\alpha_{\pm})\right)}=\frac{\Gamma\left(\frac{1}{2}(ib_{\pm}\omega-\frac{1}{2}\mp\alpha_{\pm})\right)}{\Gamma\left(\frac{1}{2}(ib_{\pm}\omega+\frac{1}{2}\mp\alpha_{\pm})\right)}\frac{{\sin}\left(\frac{\pi}{2}(-ib_{\pm}\omega+\frac{5}{2}\pm\alpha_{\pm})\right)}{{\sin}\left(\frac{\pi}{2}(-ib_{\pm}\omega+\frac{3}{2}\pm\alpha_{\pm})\right)}\nonumber\\
&~~~&~~~~~~~~~~~~~~~~~~~~~~~~~~~~~~~~\simeq-\frac{1}{\sqrt{\frac{1}{2}(ib_{\pm}\omega-\frac{1}{2}\mp\alpha_{\pm})}}\tan\left(\frac{\pi}{2}(-ib_{\pm}\omega+\frac{1}{2}\pm\alpha_{\pm})\right),\nonumber\\
&~~~&\frac{\Gamma\left(\frac{1}{2}(-ib_{\pm}\omega+\frac{3}{2}\mp\alpha_{\pm})\right)}{\Gamma\left(\frac{1}{2}(-ib_{\pm}\omega+\frac{5}{2}\mp\alpha_{\pm})\right)}=\frac{\Gamma\left(\frac{1}{2}(ib_{\pm}\omega-\frac{1}{2}\pm\alpha_{\pm})\right)}{\Gamma\left(\frac{1}{2}(ib_{\pm}\omega+\frac{1}{2}\pm\alpha_{\pm})\right)}\frac{{\sin}\left(\frac{\pi}{2}(-ib_{\pm}\omega+\frac{5}{2}\mp\alpha_{\pm})\right)}{{\sin}\left(\frac{\pi}{2}(-ib_{\pm}\omega+\frac{3}{2}\mp\alpha_{\pm})\right)}\nonumber\\
&~~~&~~~~~~~~~~~~~~~~~~~~~~~~~~~~~~~~\simeq-\frac{1}{\sqrt{\frac{1}{2}(ib_{\pm}\omega-\frac{1}{2}\pm\alpha_{\pm})}}\tan\left(\frac{\pi}{2}(-ib_{\pm}\omega+\frac{1}{2}\mp\alpha_{\pm})\right) .
\lb{Gam/Gam_x=0_1}
\eqn
To derive the above expressions, we have also made use of the properties
\bqn
\frac{\Gamma(z+m)}{\Gamma(z)}=(z)_m,~~~~\mathrm{for}~~~~\Re z>0 ,
\lb{GamRelation}
\eqn
where $m$ is a positive integer as the raising factorial is defined by Eq.~\eqref{rFactorial} and the asymptotic expansion
\bqn
\frac{\Gamma(z+\frac{1}{2})}{\Gamma(z)}\simeq\sqrt{z}\left(1-\frac{1}{8z}\right)=\sqrt{z}\left(1+{O}\left(\frac{1}{z}\right)\right) ,~~~~\mathrm{for}~~~~\Re z\to +\infty .
\lb{Gam_approximate}
\eqn
For the factor involving trigonometric functions in Eq.~\eqref{dpsi/psi_x=0_2}, we have
\bqn
\tan\left(\frac{\pi}{2}\left(-ib_{\pm}\omega+\frac{1}{2}\pm\alpha_{\pm}\right)\right)\tan\left(\frac{\pi}{2}\left(-ib_{\pm}\omega+\frac{1}{2}\mp\alpha_{\pm}\right)\right)=\frac{{\cos}(\pi\alpha_{\pm})+{\sin}(-\pi ib_{\pm}\omega)}{{\cos}(\pi\alpha_{\pm})-{\sin}(-\pi ib_{\pm}\omega)},
\lb{trigonometric_function_part}
\eqn
where one has made use of the following trigonometric relation,
\bqn
\tan A\tan B&=&\frac{\cos A\cos B+\sin A\sin B-\left(\cos A\cos B-\sin A\sin B\right)}{\cos A\cos B+\sin A\sin B+\left(\cos A\cos B-\sin A\sin B\right)}\nonumber\\
&=&\frac{\cos(A-B)-\cos(A+B)}{\cos(A-B)+\cos(A+B)} .
\lb{trigonometric_relation}
\eqn
By putting all the pieces together, the junction condition Eq.~\eqref{boundary_condition_x=0} gives
\bqn
+\sqrt{\left(ib_{+}\omega-\frac{1}{2}\right)^2-\alpha_{+}^2}\frac{{\cos}(\pi\alpha_{+})+{\sin}(-\pi ib_{+}\omega)}{{\cos}(\pi\alpha_{+})-{\sin}(-\pi ib_{+}\omega)}=-\sqrt{\left(ib_{-}\omega-\frac{1}{2}\right)^2-\alpha_{-}^2}\frac{{\cos}(\pi\alpha_{-})+{\sin}(-\pi ib_{-}\omega)}{{\cos}(\pi\alpha_{-})-{\sin}(-\pi ib_{-}\omega)}.\nb\\
\lb{dpsi/psi_x=0_3}
\eqn
which, by performing cross multiplication and using the trigonometric identities, can be reorganized and written into the following form
\begin{equation}
{\cos}\left(-2\pi i \omega b_*\right)={\cos}\left(-\pi i \omega\Delta\right)-2{\cos(\pi\alpha_{-})}{\cos(\pi\alpha_{+})}+T_\mathrm{corr}(b_{\pm},\alpha_{\pm},\omega),
\tag{\ref{high_order_result}}
\end{equation}
where $\Delta$ and $b_*$ are defined by Eq.~\eqref{Visser_parameters2}, and the correction term $T_\mathrm{corr}(b_{\pm},\alpha_{\pm},\omega)$ is given by
\begin{equation}
T_\mathrm{corr}(b_{\pm},\alpha_{\pm},\omega)=2\frac{\sqrt{\left(ib_{-}\omega-\frac{1}{2}\right)^2-\alpha_{-}^2}-\sqrt{\left(ib_{+}\omega-\frac{1}{2}\right)^2-\alpha_{+}^2}}{\sqrt{\left(ib_{-}\omega-\frac{1}{2}\right)^2-\alpha_{-}^2}+\sqrt{\left(ib_{+}\omega-\frac{1}{2}\right)^2-\alpha_{+}^2}}\left(\cos(\pi\alpha_{-})\sin(\pi ib_{+}\omega)-\cos(\pi\alpha_{+})\sin(\pi ib_{-}\omega)\right).\nb\\
\tag{\ref{correction_term}}
\end{equation}

In the remainder of this appendix, we give an account of the analytic results used in Sec~\ref{sec6}.
In the region where the effective potential is truncated $x > x_c$, we consider the following metric parameters 
\begin{equation}
(V_{0+}, b_{+}) = (0, b_{-}) ,
\tag{\ref{truncPar}}
\end{equation}
which implies
\bqn
A_{+} = B_{+}-1 &=& C_{+}-1 =-ib_{-}\omega ,\nb\\
\alpha_{+} &=& \frac12 , \lb{Euler_equation_solution_1}
\eqn
and for high overtones
\bqn
A_{\pm}, B_{\pm}, C_{\pm}< 0. \lb{Euler_equation_solution_2}
\eqn

We will focus on the asymptotical modes $|\Im\omega| \gg \Re\omega$.
To assess the junction condition, we first evaluate the Taylor expansions about $x_0$ to approximate the ratio between the first-order derivative of the wave functions $\partial_{x}\psi_{\pm}$ and the wave functions $\psi_{\pm}$.
Specifically, $x_c=x_0+\Delta x$, where $x_0=0$.
We have
\bqn
\left.\frac{\partial_{x}\psi_{\pm}(\omega,x)}{\psi_{\pm}(\omega,x)}\right|_{x_c}&=&\left.\left[\frac{\partial_{x}\psi_{\pm}(\omega,x)}{\psi_{\pm}(\omega,x)}+\partial_{x}\frac{\partial_{x}\psi_{\pm}(\omega,x)}{\psi_{\pm}(\omega,x)}\Delta x+\frac{1}{2}\partial_{xx}\frac{\partial_{x}\psi_{\pm}(\omega,x)}{\psi_{\pm}(\omega,x)}(\Delta x)^2+\cdots+{O}((\Delta x)^n)\right]\right|_{x_0}\nonumber\\
&=&\bigg[\frac{\partial_{x}\psi_{\pm}(\omega,x)}{\psi_{\pm}(\omega,x)}+\left(\frac{\partial_{xx}\psi_{\pm}(\omega,x)\psi_{\pm}(\omega,x)-(\partial_{x}\psi_{\pm}(\omega,x))^2}{(\psi_{\pm}(\omega,x))^2}\right)\Delta x\nonumber\\
&~&+\frac{1}{2}\bigg(\frac{\partial_{xxx}\psi_{\pm}(\omega,x)\psi_{\pm}(\omega,x)-\partial_{xx}\psi_{\pm}(\omega,x)\partial_{x}\psi_{\pm}(\omega,x)}{(\psi_{\pm}(\omega,x))^2}\nonumber\\
&~&-2\frac{\partial_{x}\psi_{\pm}(\omega,x)}{\psi_{\pm}(\omega,x)}\left(\frac{\partial_{xx}\psi_{\pm}(\omega,x)\psi_{\pm}(\omega,x)-(\partial_{x}\psi_{\pm}(\omega,x))^2}{(\psi_{\pm}(\omega,x))^2}\right)\bigg)(\Delta x)^2+\cdots+{O}((\Delta x)^n)\bigg]\bigg|_{x_0}.\nb\\
\lb{dpsi/psi_x0=x1-Deltax}
\eqn

To gain some insights, let us revisit the derivations in the main text and consider the first few terms in the expansion.
Using Eqs.~\eqref{discontinuous_PT_wavefunction} and~\eqref{differential_discontinuous_PT_wavefunction}, the first term on the r.h.s. of Eq.~\eqref{dpsi/psi_x0=x1-Deltax} gives
\bqn
\left.\frac{\partial_{x}\psi_{\pm}(\omega,x)}{\psi_{\pm}(\omega,x)}\right|_{x_0}&=&\mp\frac{1}{2}\frac{(A_{\pm})(B_{\pm})}{(C_{\pm})}\frac{{_{2}F_1}(A_{\pm}+1,B_{\pm}+1,C_{\pm}+1,\frac{1}{2})}{{_{2}F_1}(A_{\pm},B_{\pm},C_{\pm},\frac{1}{2})}\nonumber\\
&=&\mp\frac{1}{2}\frac{(A_{\pm})(B_{\pm})}{(C_{\pm})}\frac{\Gamma(C_{\pm}+1)}{\Gamma(C_{\pm})}\frac{\Gamma\left(\frac{1}{2}(A_{\pm}+1)\right)}{\Gamma\left(\frac{1}{2}(A_{\pm}+2)\right)}\frac{\Gamma\left(\frac{1}{2}(B_{\pm}+1)\right)}{\Gamma\left(\frac{1}{2}(B_{\pm}+2)\right)}\nonumber\\
&=&\mp2\frac{\Gamma\left(\frac{1}{2}(A_{\pm}+1)\right)}{\Gamma\left(\frac{1}{2}A_{\pm}\right)}\frac{\Gamma\left(\frac{1}{2}(B_{\pm}+1)\right)}{\Gamma\left(\frac{1}{2}B_{\pm}\right)}\nonumber\\
&=&\mp2\frac{\Gamma\left(\frac{1}{2}(1-A_{\pm})+\frac{1}{2}\right)}{\Gamma\left(\frac{1}{2}(1-A_{\pm})\right)}\frac{\Gamma\left(\frac{1}{2}(1-B_{\pm})+\frac{1}{2}\right)}{\Gamma\left(\frac{1}{2}(1-B_{\pm})\right)}\tan(\frac{\pi}{2}A_{\pm})\tan(\frac{\pi}{2}B_{\pm})\nonumber\\
&\simeq&\mp\sqrt{(1-A_{\pm})(1-B_{\pm})}\tan\left(\frac{\pi}{2}A_{\pm}\right)\tan\left(\frac{\pi}{2}B_{\pm}\right).
\lb{dpsi/psi_x0_0}
\eqn
On the l.h.s. of the truncation point, it simplifies to
\bqn
\left.\frac{\partial_{x}\psi_{-}(\omega,x)}{\psi_{-}(\omega,x)}\right|_{x_0}&\simeq&+\sqrt{\left(ib_{-}\omega-\frac{1}{2}\right)^2-\alpha_{-}^2}\frac{{\cos}(\pi\alpha_{-})+{\sin}(-\pi ib_{-}\omega)}{{\cos}(\pi\alpha_{-})-{\sin}(-\pi ib_{-}\omega)}\simeq ib_{-}\omega\frac{{\cos}(\pi\alpha_{-})+{\sin}(-\pi ib_{-}\omega)}{{\cos}(\pi\alpha_{-})-{\sin}(-\pi ib_{-}\omega)},\nb\\
\lb{dpsi/psi_x0_lhs}
\eqn
where one makes use of Eq.~\eqref{trigonometric_relation}.
While on the r.h.s. of the truncation point, Eq.~\eqref{dpsi/psi_x0_0} gives
\bqn
\left.\frac{\partial_{x}\psi_{+}(\omega,x)}{\psi_{+}(\omega,x)}\right|_{x_0}&\simeq&-\sqrt{(ib_{-}\omega+1)(ib_{-}\omega)}\tan\left(-\frac{\pi}{2}ib_{-}\omega\right)\tan\left(-\frac{\pi}{2}ib_{-}\omega+\frac{\pi}{2}\right)
=\sqrt{(ib_{-}\omega+1)(ib_{-}\omega)}
\simeq ib_{-}\omega.\nb\\
\lb{dpsi/psi_x0_rhs}
\eqn
By equating Eq.~\eqref{dpsi/psi_x0_lhs} to~\eqref{dpsi/psi_x0_rhs}, one immediately finds the following purely imaginary solutions
\begin{equation}
\omega=-i\frac{n}{b_{-}} , \tag{\ref{dpsi/psi_x0_final_2}}
\end{equation}
when $\sin(-\pi ib_{-}\omega)=0$.

For the second term on the r.h.s. of Eq.~\eqref{dpsi/psi_x0=x1-Deltax}, one observes the coefficient given by
\bqn
\left.\frac{\partial_{xx}\psi_{\pm}(\omega,x)\psi_{\pm}(\omega,x)-(\partial_{x}\psi_{\pm}(\omega,x))^2}{(\psi_{\pm}(\omega,x))^2}\right|_{x_0}=\left.\left(\frac{\partial_{xx}\psi_{\pm}(\omega,x)}{\psi_{\pm}(\omega,x)}-\left(\frac{\partial_{x}\psi_{\pm}(\omega,x)}{\psi_{\pm}(\omega,x)}\right)^2\right)\right|_{x_0} .
\lb{dpsi/psi_x0_1}
\eqn
The only novelty comes from the ratio between the waveform's second-order derivative to the waveform, namely,
\bqn
\left.\frac{\partial_{xx}\psi_{\pm}(\omega,x)}{\psi_{\pm}(\omega,x)}\right|_{x_0}&=&\left(\mp\frac{1}{2}\right)^2\frac{(A_{\pm})_{2}(B_{\pm})_{2}}{(C_{\pm})_{2}}\frac{{_{2}F_1}(A_{\pm}+2,B_{\pm}+2,C_{\pm}+2,\frac{1}{2})}{{_{2}F_1}(A_{\pm},B_{\pm},C_{\pm},\frac{1}{2})}\nonumber\\
&=&\left(\mp\frac{1}{2}\right)^2\frac{(A_{\pm})_{2}(B_{\pm})_{2}}{(C_{\pm})_{2}}\frac{\Gamma(C_{\pm}+2)}{\Gamma(C_{\pm})}\frac{\Gamma\left(\frac{1}{2}(A_{\pm}+1)\right)}{\Gamma\left(\frac{1}{2}(A_{\pm}+3)\right)}\frac{\Gamma\left(\frac{1}{2}(B_{\pm}+1)\right)}{\Gamma\left(\frac{1}{2}(B_{\pm}+3)\right)}\nonumber\\
&=&(A_{\pm})(B_{\pm}) .
\lb{dpsi/psi_x0_1_1}
\eqn
For asymptotical modes, this term is identical for waveforms on both sides of the truncation point.

Similarly, for the third term on the r.h.s. of Eq.~\eqref{dpsi/psi_x0=x1-Deltax}, one only needs to pay attention to the ratio
\bqn
\left.\frac{\partial_{xxx}\psi_{\pm}(\omega,x)}{\psi_{\pm}(\omega,x)}\right|_{x_0}&=&\left(\mp\frac{1}{2}\right)^3\frac{(A_{\pm})_{3}(B_{\pm})_{3}}{(C_{\pm})_{3}}\frac{{_{2}F_1}(A_{\pm}+3,B_{\pm}+3,C_{\pm}+3,\frac{1}{2})}{{_{2}F_1}(A_{\pm},B_{\pm},C_{\pm},\frac{1}{2})}\nonumber\\
&=&\left(\mp\frac{1}{2}\right)^3\frac{(A_{\pm})_{3}(B_{\pm})_{3}}{(C_{\pm})_{3}}\frac{\Gamma(C_{\pm}+3)}{\Gamma(C_{\pm})}\frac{\Gamma\left(\frac{1}{2}(A_{\pm}+1)\right)}{\Gamma\left(\frac{1}{2}(A_{\pm}+4)\right)}\frac{\Gamma\left(\frac{1}{2}(B_{\pm}+1)\right)}{\Gamma\left(\frac{1}{2}(B_{\pm}+4)\right)}\nonumber\\
&=&\mp2(A_{\pm}+1)(B_{\pm}+1)\frac{\Gamma\left(\frac{1}{2}(A_{\pm}+1)\right)}{\Gamma\left(\frac{1}{2}A_{\pm}\right)}\frac{\Gamma\left(\frac{1}{2}(B_{\pm}+1)\right)}{\Gamma\left(\frac{1}{2}B_{\pm}\right)}\nonumber\\
&=&\mp2(A_{\pm}+1)(B_{\pm}+1)\frac{\Gamma\left(\frac{1}{2}(1-A_{\pm})+\frac{1}{2}\right)}{\Gamma\left(\frac{1}{2}(1-A_{\pm})\right)}\frac{\Gamma\left(\frac{1}{2}(1-B_{\pm})+\frac{1}{2}\right)}{\Gamma\left(\frac{1}{2}(1-B_{\pm})\right)}\tan(\frac{\pi}{2}A_{\pm})\tan(\frac{\pi}{2}B_{\pm})\nonumber\\
&\simeq&\mp(A_{\pm}+1)(B_{\pm}+1)\sqrt{(1-A_{\pm})(1-B_{\pm})}\tan\left(\frac{\pi}{2}A_{\pm}\right)\tan\left(\frac{\pi}{2}B_{\pm}\right).
\lb{dpsi/psi_x0_1_2}
\eqn
Again, by employing very similar arguments regarding Eqs.~\eqref{dpsi/psi_x0_lhs} and~\eqref{dpsi/psi_x0_rhs}, it can be shown that Eq.~\eqref{dpsi/psi_x0_final_2} are the solutions for asymptotic modes.

Therefore, to generalize the above results, it suffices to show that the ratio for an arbitrary order
\bqn
\left.\frac{\partial_{\underbrace{x\cdots x}_n}\psi_{\pm}(\omega,x)}{\psi_{\pm}(\omega,x)}\right|_{x_0} ,\nb
\eqn
contains factors of $\tan\left(\frac{\pi}{2}A_{\pm}\right)\tan\left(\frac{\pi}{2}B_{\pm}\right)$, apart from the remaining irrelevant part which is identical on both sides of the truncation for asymptotic modes.
The above statement can be manifestly shown by noticing that the ratio can be generally rewritten as
\bqn
\left.\frac{\partial_{\underbrace{x\cdots x}_n}\psi_{\pm}(\omega,x)}{\psi_{\pm}(\omega,x)}\right|_{x_0} 
&=&\left(\mp\frac{1}{2}\right)^n\frac{(A_{\pm})_{n}(B_{\pm})_{n}}{(C_{\pm})_{n}}\frac{{_{2}F_1}(A_{\pm}+n,B_{\pm}+n,C_{\pm}+n,\frac{1}{2})}{{_{2}F_1}(A_{\pm},B_{\pm},C_{\pm},\frac{1}{2})}\nonumber\\
&=& \left(\mp\frac{1}{2}\right)^n(A_{\pm})_{n}(B_{\pm})_{n}\frac{\Gamma\left(\frac{1}{2}(A_{\pm}+1)\right)}{\Gamma\left(\frac{1}{2}(A_{\pm}+n+1)\right)}\frac{\Gamma\left(\frac{1}{2}(B_{\pm}+1)\right)}{\Gamma\left(\frac{1}{2}(B_{\pm}+n+1)\right)}\nonumber\\
&=&\left\{
    \begin{aligned}
    \mp2(A_{\pm}+1)\cdots(A_{\pm}+(n-2))(B_{\pm}+1)\cdots(B_{\pm}+(n-2))~~~~~~~~~~~~~~\\
    \frac{\Gamma\left(\frac{1}{2}(A_{\pm}+1)\right)}{\Gamma\left(\frac{1}{2}A_{\pm}\right)}\frac{\Gamma\left(\frac{1}{2}(B_{\pm}+1)\right)}{\Gamma\left(\frac{1}{2}B_{\pm}\right)}~~~~~\mathrm{for~~~~odd}~~~n\geq3\\\\
    (A_{\pm})\cdots(A_{\pm}+(n-2))(B_{\pm})\cdots(B_{\pm}+(n-2))~~~~~\mathrm{for~~~~even}~~n\geq2\\
    \end{aligned}
\right.,
\eqn
which encloses the proof.

\bibliographystyle{JHEP}
\bibliography{references_qian}

\end{document}